# Transitional and Near-Wall Turbulence Dynamics over Rib-Roughened Surfaces

Ranjan Kushwaha[1], S. Sarkar[1], Gautam Biswas [1, 2 *]

[1]Department of Mechanical Engineering, Indian Institute of Technology, Kanpur, Uttar Pradesh-208016, India

[2]Department of Mechanical Engineering, Birla Institute of Technology and Science, Pilani, K.K. Birla Goa Campus, Sancoale, Goa 403726, India

* Email: gtm@iitk.ac.in

This study utilizes Large Eddy Simulation (LES) to investigate the impact of longitudinal triangular riblets on the laminar-to-turbulent transition in boundary layer flow. Five cases are examined: one involving a flat plate and four with ribbed plates. Among the ribbed cases, three use a riblet aspect ratio of two, whereas one has an aspect ratio of one. Arrays of longitudinal triangular riblets are positioned on a flat plate, and the transition to turbulence is initiated by controlled excitation of a Tollmien–Schlichting (TS) wave, imposed on a Blasius velocity profile in stable region. The longitudinal triangular riblets attenuate the TS wave, leading to a lower growth rate of turbulence. For higher riblet height ($h$) and width ($w$) (inner-scaled dimensions are $h^+ = 25, w^+ = S^+ = 50$, $S$ is the spacing between two riblets), an early transition is triggered by high-frequency disturbances generated at the leading edge of the roughness elements, though increasing riblet spacing ($S^+ = 75$) delays the transition by 17.5%. Both cases exhibited increased drag compared to the flat plate. For $h^+ = 12.5$ and $w^+ = S^+ = 25$, transition was delayed by 37%, with a modest shear-drag reduction of 8.8%. The most significant result from considered cases, $h^+ = w^+ = S^+ = 12.5$, showed a 47% delay in transition and a 13.69% reduction in shear-drag. Smaller riblets cause minimal disturbance at the leading edge of roughness, resulting in a transition mechanism similar to a flat plate, while also reducing pressure loss, secondary flows, and velocity fluctuations.

## I. INTRODUCTION

Organized structures in turbulent flows significantly impact turbulent transport (Cantwell [1]; Robinson [2]). Despite their randomness, turbulent flows exhibit coherent structures such as quasi-streamwise vortices, hairpin vortices, and hairpin packets, which persist in wall turbulence (Adrian et al. [3]). These structures show correlated scales much larger than the smallest flow scales and are characterized by significant correlations in fundamental variables over extended spatial and temporal ranges (Saha et al. [4]). In natural flow transition, spanwise vortices transform into lambda vortices due to vortex filament redirection and reorganization, forming counter-rotating legs with opposite streamwise vorticity. Hairpin vortices then develop, with a head, neck, and legs. Subsequent vortex rings form separately due to shear layer instability (K-H instability), advancing turbulence development [5–7].

Surface-mounted riblets effectively reduce drag in turbulent boundary layers despite increasing wetted area. Triangular riblets achieve optimal drag reduction of 7–8% at a spacing ($S^+$) of ~15 (Walsh and Linderman [8]). Consistent reductions occur with groove heights ($h^+$) below 18 and spacing below 25 wall units (Coustols and Coustex [9]; Walsh and Weinstein [10]). Riblets suppress near-wall turbulence, reducing quasi-streamwise vortices and skin-friction drag (Choi et al. [11]; Sirovich and Karlsson [12]). The impact of riblets on drag reduction has been quantified by various methods (Bechert and Bartenwerfer [13]; Luchini, Manzo and Pozzi [14]). Skin-friction reduction persists under different adverse pressure gradients (Nieuwstadt et al. [15]). Various methods have been explored for reducing skin friction, such as riblets, large-eddy breakup devices, polymer additives, and compliant surfaces (Bushnell and McGinley [16]; Coustols and Savill [17]). Riblet's effectiveness peaks and then degrades with increasing size due to secondary vortices and flow disruption (Goldstein and Tuan [18]; García-Mayoral and Jiménez [19]). Turbulence shifts and drag reduction have been further quantified (García-Mayoral et al. [20]; Orlandi and Leonardi [21]). Typically, surface roughness increases skin-friction drag, degrading performance in systems like biofouled ship hulls (Schultz et al. [22]). Riblets maintain effectiveness under adverse pressure gradients (Nieuwstadt et al. [15]) and can lower drag by up to 10% (Bechert et al. [23]). The effect of riblets on the laminar-turbulent transition has been explored in several studies, leading to mixed conclusions and highlighting the need for further research. In the context of fully turbulent boundary layers, Walsh and Weinstein [9] suggested that riblets should be designed to confine turbulent bursts to their origin, minimizing their transverse spread. Belov, Enutin, and Litvinov [24] did not find any favorable impact of riblets on the laminar-turbulent transition, regardless of whether riblets were aligned with or across the flow. Their study involved placing riblets on the entire surface of the flat plate where the transition occurred. Kozlov et al. [25, 26] found that longitudinal grooves on a test plate could either enhance or inhibit bypass transition in the presence of high free-stream turbulence, depending on factors such as pressure gradient and leading-edge shape. Neumann and Dinkelacker [27] discovered that riblets delay the formation of initial turbulent structures in the transitional flow zone on a body of V-grooved surfaces. Chu, Henderson, and Karniadakis [28] used numerical simulations to show that riblets can reduce drag during the transitional regime compared with a smooth surface. Luchini [29] suggested that riblets can destabilize Tollmien–Schlichting (TS) waves. Grek et al. [30] focused on the influence of riblets on the development of TS waves and Λ-vortex breakdown without explicitly addressing the adverse effects of boundary-layer tripping caused by riblets. Klumpp et al. [31] demonstrated the effectiveness of scalloped riblets in controlling the transition of zero-pressure-gradient boundary layers under various forced transition scenarios using large-eddy simulations. Wang et al. [32] conducted direct numerical simulations (DNS) to study the effect of riblet tip sharpness on drag reduction. They found that scalloped riblets with slightly curved tips are more effective in delaying transition and reducing drag compared to sharp-tipped riblets.

The conflicting findings in previous studies highlight the challenges in understanding the complex mechanisms behind laminar-to-turbulent transition, particularly in relation to longitudinal triangular riblets. This study aims to address these gaps by focusing on the development of disturbances, such as the TS waves in stable regions, at both linear and nonlinear stages of transition. By carefully controlling disturbances within a laminar boundary layer, we aim to gain deeper insights into the influence of riblets with varying heights, widths, and spacings on the transition process. The primary objective is to understand how different riblet dimensions affect the disturbances involved in transition, offering a potential explanation for their role in both delaying transition and reducing drag. Additionally, the study will assess the overall pressure loss of fluids and drag reduction across the entire plate, with a specific focus on analyzing the turbulent kinetic energy (TKE) budget and the secondary flow to better understand near-wall turbulence behavior during the transition. This study marks an important step towards clarifying the role of riblets in transition control and turbulence management.

## II. FORMULATION AND METHOD

This section outlines the computational methodology and the geometry of interest. In Large Eddy Simulation (LES), the primary focus is on accurately simulating the dynamics of large, energy-containing structures, which play a crucial role in the transport of momentum and energy. To account for the influence of smaller, unresolved scales, subgrid-scale (SGS) modelling is employed. These smaller scales tend to exhibit more isotropic and consistent behaviour, with a relatively smaller dynamic impact. In the governing equations, the influence of these unresolved scales on the resolved scales is represented by a residual stress term. This term is derived from the grid-scale velocity field, allowing for the closure of the governing equations. A filtering operation is applied to the incompressible Navier–Stokes and continuity equations, leading to filtered equations of motion that describe the evolution of the large, energy-carrying eddies. These filtered equations are typically expressed in tensor notation, as shown below:

$$\frac{\partial \bar{u}_i}{\partial x_i} = 0 \tag{1}$$

$$\frac{\partial \bar{u}_i}{\partial t} + \frac{\partial \bar{u}_i \bar{u}_j}{\partial x_j} = -\frac{\partial \bar{p}}{\partial x_i} - \frac{\partial \tau_{ij}}{\partial x_j} + \frac{1}{Re}\frac{\partial^2 \bar{u}_i}{\partial x_j \partial x_j} \tag{2}$$

In the above equations, the indices $i$ and $j$ (where $i, j$ = 1, 2, 3) correspond to the spatial coordinates $x$ (streamwise), $y$ (cross-stream), and $z$ (spanwise). In this formulation, $\bar{u}_i$ represents the instantaneous filtered velocity in the $i$ direction, $\bar{p}$ signifies the filtered pressure, and $t$ denotes the time variable. The velocity and length scales in the above equations are normalized by freestream velocity ($U_\infty$) and inlet displacement thickness ($\delta_{in}^*$) at the inlet, respectively. The Reynolds number based on the displacement thickness at the inlet is defined as $Re = (U_\infty\, \delta^*)/\nu$, where $\nu$ is the kinematic viscosity. The term $\tau_{ij}$ in Eq. (2) symbolizes the influence of small-scale eddies in the large-scale transport equation. The subgrid-scale model introduced by Smagorinsky (33) relies on the gradient transport hypothesis. This hypothesis establishes a correlation between $\tau_{ij}$ and the large-scale strain rate tensor $\bar{S}_{ij}$. The model posits that the subgrid-scale stress tensor is proportional to the local strain rate, which is given as,

$$\tau_{ij} - \frac{\delta_{ij}}{3}\tau_{kk} = -2\nu_T \bar{S}_{ij} \tag{3}$$

Here $\nu_T$= eddy viscosity; $\delta_{ij}$ = Kronecker delta, and

$$\bar{S}_{ij} = \frac{1}{2}\left(\frac{\partial \bar{u}_i}{\partial x_j} + \frac{\partial \bar{u}_j}{\partial x_i}\right) \tag{4}$$

Lilly [34] proposed the following formula to obtain the eddy viscosity as:

$$\nu_T = (C_s \bar{\Delta})^2 |\bar{S}| \tag{5}$$

Here $|\bar{S}| = \sqrt{\left(2\bar{S}_{ij}\bar{S}_{ij}\right)}$ and $C_s$ is the Smagorinsky constant, $\bar{\Delta}$ is the grid filter scale. Substitution of Eq. (5) in Eq. (3) yields,

$$\tau_{ij} - \frac{\delta_{ij}}{3}\tau_{kk} = -2C\beta_{ij} \tag{6}$$

With

$$\beta_{ij} = \bar{\Delta}^2 |\bar{S}| \bar{S}_{ij} \; ; \; C = C_s^2 \tag{7}$$

Germano et al. [35] introduced a dynamic procedure for calculating the eddy viscosity coefficient $C$ at each time step and grid point based on the grid filtered velocity field. This approach involves the utilization of both, a grid filter (denoted by an overbar) representing resolved and subgrid scales, and a test filter (indicated by a caret over the overbar) with a width greater than that of the grid filter.

The test filter defines a new set of stresses, specifically, the subtest-scale stresses $T_{ij}$. Mathematically, these subtest-scale stresses can be expressed as following:

$$T_{ij} = \overline{\widehat{u_i u_j}} - \hat{\bar{u}}_i \hat{\bar{u}}_j \tag{8}$$

Where, $\hat{\bar{u}}_i$ is the test filtered velocity field which is calculated by a procedure of Najjar and Tafti [36] on a test level grid-mesh.

Test level subgrid-scale stresses can also be expressed in the terms of Smagorinsky closure as

$$T_{ij} - \frac{\delta_{ij}}{3} T_{kk} = -2C\alpha_{ij} \tag{9}$$

With

$$\alpha_{ij} = \hat{\bar{\Delta}}^2 \left| \hat{\bar{S}} \right| \hat{\bar{S}}_{ij} \tag{10}$$

where,

$$\hat{\bar{S}}_{ij} = \frac{1}{2} \left( \frac{\partial \hat{\bar{u}}_i}{\partial x_j} + \frac{\partial \hat{\bar{u}}_j}{\partial x_i} \right) \tag{11}$$

And

$$\hat{\bar{\Delta}} = \left( \hat{\bar{\Delta}}_1 \, \hat{\bar{\Delta}}_2 \, \hat{\bar{\Delta}}_3 \right)^{\frac{1}{3}}, \frac{\hat{\bar{\Delta}}}{\bar{\Delta}} = 2 \tag{12}$$

Here $\hat{\bar{\Delta}}$ is the test level filter scale.

The significant advancement introduced by Germano et al. [34] in subgrid-scale modelling lies in the identification that the consistency between Eq. (6) and (8) relies on a judicious selection of the eddy viscosity coefficient. The crux of this achievement involves subtracting the test-scale average of the resolved and subgrid-scale stresses $\tau_{ij}$ from the test-scale stresses $T_{ij}$. Mathematically, this subtraction can be expressed as (Najjar and Tafti [36])

$$L_{ij} = l_{ij} - \frac{\delta_{ij}}{3} l_{kk} = T_{ij} - \hat{\tau}_{ij} = -2C\alpha_{ij} + \widehat{2C\beta_{ij}} \tag{13}$$

It is difficult to calculate the value of $C$ since it appears within a filter operation. Piomelli and Liu [37] proposed a more straightforward approach by modifying Eq. (13) as,

$$L_{ij} = -2C\alpha_{ij} + 2\widehat{C^*\beta_{ij}} \tag{14}$$

Specifically, on the right-hand side of Eq. (14), an estimate of the coefficient ($C$) is replaced by $C^*$. The value of $C^*$ is assumed to be known. In this scenario, the minimization of the sum of the squares is carried out, resulting in an optimized solution given as,

$$C^n(x, y, z) = -\frac{1}{2} \frac{\left(L_{ij} - 2C^*\widehat{\beta_{ij}}\right)\alpha_{ij}}{\alpha_{mn}\alpha_{mn}} \tag{15}$$

Piomelli and Liu [37] suggested that the value of $C^*$ can be taken as previous timestep value of $C$ i.e. $C^* = C^{n-1}$. Superscript $n-1$ represents the values at previous time level. In the current calculation, despite employing the local averaging method introduced by Zang et al. [38] spurious values of parameter $C$ emerged. Consequently, to address this issue and prevent the occurrence of negative value of $C$, an additional constraint was applied to the averaged value of $C$, that means $C$ must be greater than or equal to zero (Ghosal et al. [39]). The clipping of the dynamic coefficient to non-negative values was implemented to ensure numerical stability and avoid non-physical negative eddy viscosities. The spatial discretization uses a second-order central difference scheme and temporal evolution employs the second-order accurate Adams-Bashforth scheme [40]. The Poisson equation for pressure correction is solved using the Bi-CGSTAB method, which is an iterative method to solve the system of linear equations, developed by van der Vorst [41]. A detailed description of the dynamic subgrid stress model and the solution algorithm incorporated in this work is available elsewhere [42].

The computational domain is shown in Figure 1 for the ribbed plate case. Apart from the flat plate case (FP), four ribbed plate cases are considered with different widths, heights, and spacing of the riblet, as shown in Table 1. Among the ribbed cases, three used a riblet aspect ratio of 2, whereas one has an aspect ratio of 1. The domain size is taken as $490\delta_{in}^*$ in $x$ direction, $15\delta_{in}^*$ in $y$ direction, and $30\delta_{in}^*$ in $z$ direction [43]. For the ribbed plate cases, the longitudinal riblets are placed at the wall in an axial direction that starts from $x = 35$ to the end of the domain.

$$v'(x, y = 0, z, t) = a_f \, exp[-b_f \left(x - c_f\right)^2] \, sin(\omega t) \, sin(\beta z) \quad (16)$$

The equation of the applied disturbance strip [46] is written in Eq.16. Here $af$, $b_f$, and $c_f$ are the constants controlling the streamwise variation of the forcing, $\omega$ is the frequency, and $\beta$ is the spanwise wavenumber. The value of $b_f$, $c_f$, and $\beta$ is taken as 0.125, 10, and 0.41, respectively [46]. The value for $af$ and $\omega$ is discussed in the next section. The uniform grid resolution, $\delta x^+ \approx 16.33$, $\delta z^+ \leq 2.23$ ($\delta x^+ = u_\tau \delta x/\nu$), and non-uniform grid in $y$ direction with $\delta y^+$ based on the first grid point is close to unity or below unity, which is sufficient to resolve the laminar sublayer. Employed grid resolutions is sufficient to capture the transition and fully developed turbulence regions accurately [43-45]. In Table 1, $S$ is the spacing between riblets, $w$ is the width of riblets, and $h$ is the height of riblets. The number of grid points in the streamwise ($N_x$), wall-normal ($N_y$), and spanwise ($N_z$) directions is summarized in Table 1. The total number of cells in the computational domain exceeds 18.5 million (For RP4, the mesh consists of approximately 37 million grid cells). To assess grid independence for both the FP and RP1 cases, results were compared between three simulations using 8 million to 18.5 million grid cells. The change in the root mean square (rms) values of velocity fluctuations between the 12 million and 18.5 million cell cases was found to be less than 2% in the fully developed turbulence regions. In the transitional regions, the predicted locations of transition onset varied by no more than 1.8% as compare to FP cases with 18.5 million grid pints. For the FP case, the transition location was identified using simulations with 12 million and 18.5 million grid points shown in the Figure 1(b). Following the methodology proposed by Ol et al. [52] and Jain and Sarkar [43], the onset of transition was determined based on a critical Reynolds shear stress value of 0.001. These results confirm that the grid resolution employed in the present study is sufficient to accurately capture the relevant flow features. Periodic boundary conditions have been applied to ensure continuity in spanwise directions. At the inlet plane, the Blasius profile is imposed for $u$ corresponding to Reynolds number ($Re = \frac{U_\infty \delta_{in}^*}{\nu}$) of 500, whereas cross-stream velocities $v$ and $w$ are assumed to be zero. A convective boundary condition is imposed at the exit of the domain. The lower boundary enforces a no-slip condition ( $u = v = w = 0$), whereas the upper boundary adopts a free-slip ($\partial u/\partial y = \partial w/\partial y = v = 0$) condition. The computational parallelization is implemented through Open-MP.

Table 1. Details of various computational domains used in the present study.

| Nomenclature | $L_x$, $L_y$, $L_z$ | $h^+$, $w^+$, $S^+$ | $N_x$, $N_y$, $N_z$ | $\delta x^+, \delta y^+, \delta z^+$ | $Re = \frac{U_\infty \delta_{in}^*}{\nu}$ |
|---|---|---|---|---|---|
| FP (flat plate) | 490, 15, 30 | -------- | 752, 82, 302 | 16.33, 0.625-6.25, 2.23 | 500 |
| RP1 | 490, 15, 30 | 25, 50, 50 | 752, 82, 302 | 16.33, ≈ 1.0-7.0, 2.23 | 500 |
| RP2 | 490, 15, 30 | 25, 50, 75 | 752, 82, 302 | 16.33, ≈ 1.0-7.0, 2.23 | 500 |
| RP3 | 490, 15, 30 | 12.5, 25, 25 | 752, 82, 302 | 16.33, ≈ 1.0-7.0, 2.23 | 500 |
| RP4 | 490, 15, 30 | 12.5, 12.5, 12.5 | 752, 82, 602 | 16.33, ≈ 1.0-7.0, 1.12 | 500 |

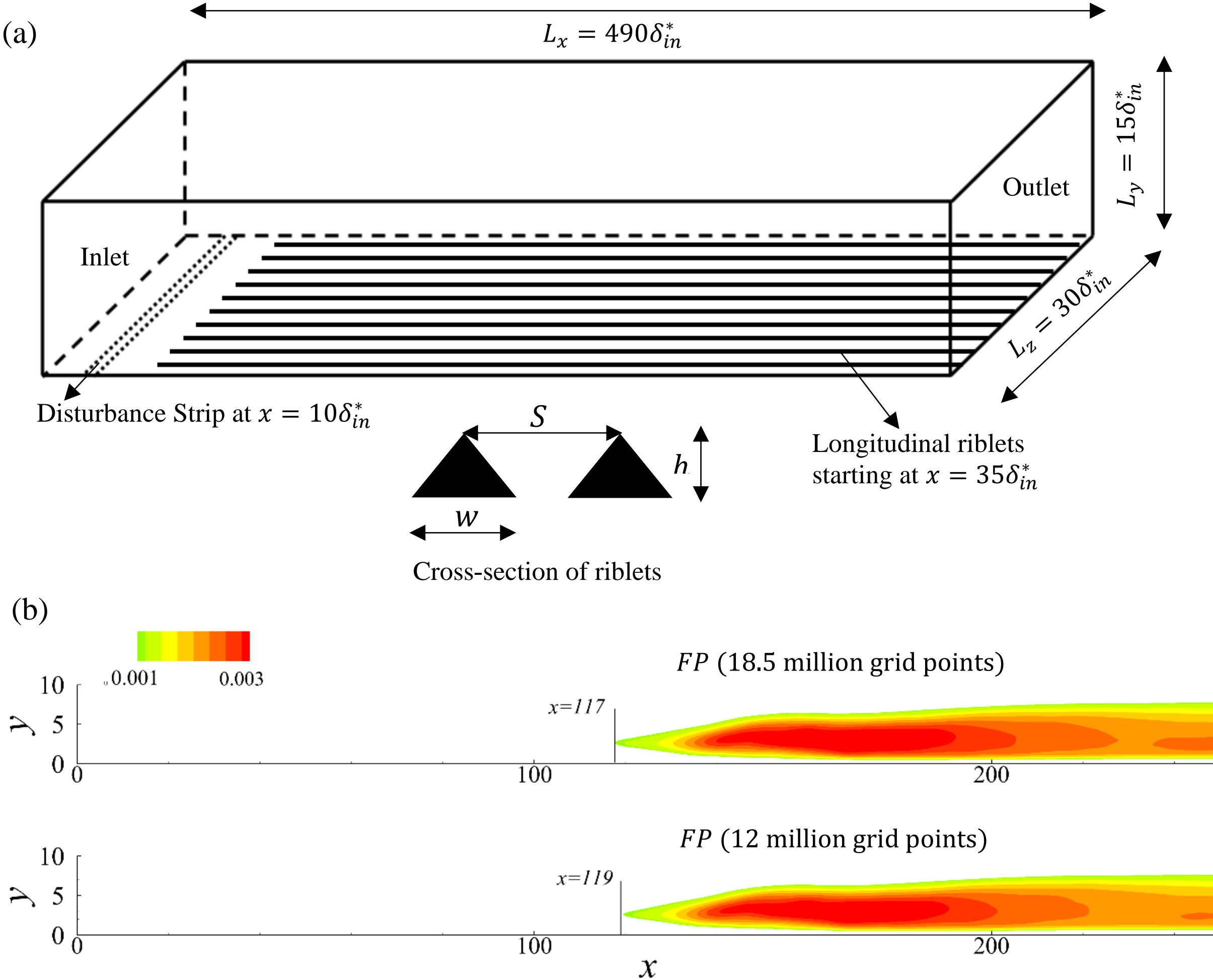

Figure 1. (a) Computational domain for the ribbed plate cases. The ribs are placed at the wall in axial direction that starts from $x = 35\delta_{in}^*$ to the end of the domain. Here, $S$ is the spacing between riblets, $w$ is the width of riblets, and $h$ is the height of riblets; (b) Contour of spanwise averaged Reynolds shear stress ($-\overline{u'v'}$) for FP case with 18.5 and 12 million grid cells.

We may mention that, in the above formulation, the grid filtered quantities are indicated by an overbar. However, in the rest of the paper, an overbar will signify average quantities, while without an overbar, it will represent instantaneous (grid filtered) quantities.

## III. VALIDATION

For the present study, the Renolds number based on the friction velocity and the boundary layer thickness ($Re_\tau = (u_\tau\ \delta)/\nu$) is found to approximately 175 for the flat plate (FP) case in the fully developed turbulent flow regime. Hence, our simulation results are validated against Choi's [11] DNS results for a same size of periodic channel flow configuration with frictional Renolds number ($Re_\tau$) of 180. In this configuration, the upper wall is a flat plate, whereas the lower wall features triangular riblets ($w^+ = 2h^+ = S^+$ with $h^+ = 20$). Periodic boundary conditions have been applied in both axial and spanwise directions. The lower and upper boundary enforces a no-slip boundary condition. The uniform grid resolution is taken for the validations are, $\delta x^+ = 17.88$, $\delta z^+$= 2.5 and non-uniform grid in $y$ direction with $\delta y^+$ based on the first grid point is close to unity. Figure 2 shows the validation of the axial velocity profile normalised with centreline velocity of a laminar parabolic profile with the same volume flux and root mean square value, which is normalised with centreline velocity, which closely matches the DNS results reported by Choi [11]. For a comprehensive validation of the present in-house LES code, the reader is referred to our previous publications [42, 44, 45].

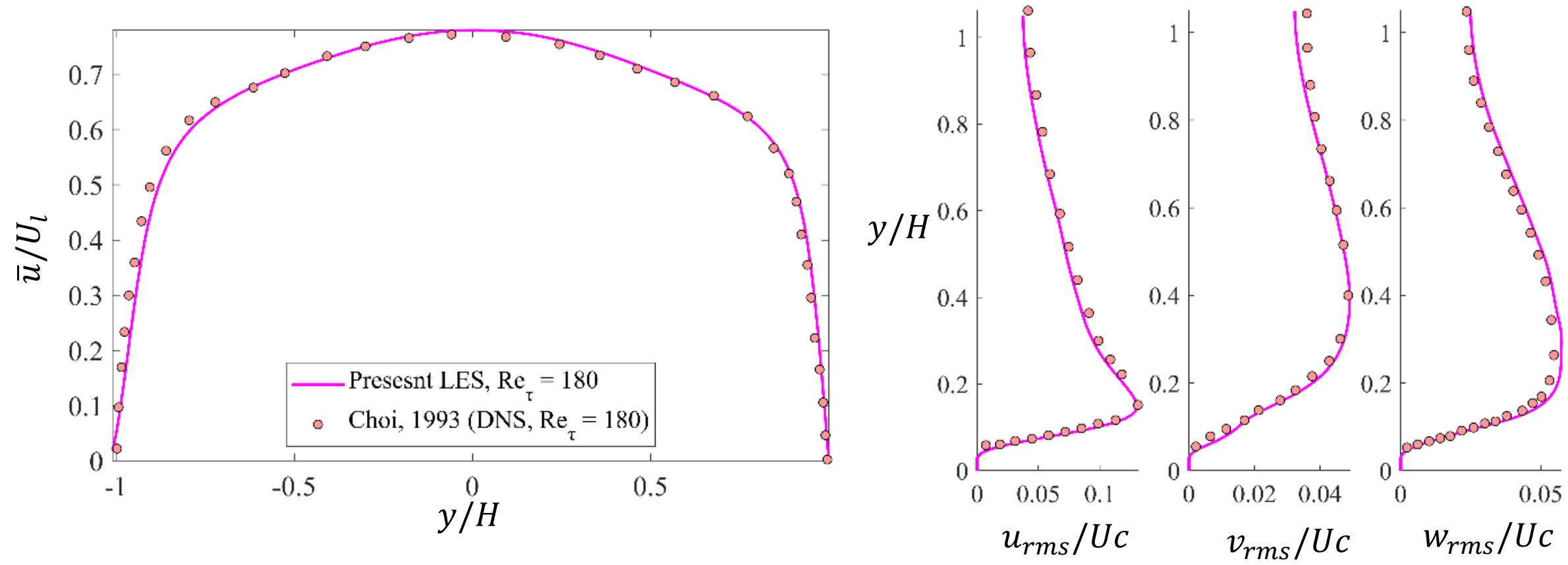

Figure 2. Validation of axial velocity profile and root mean square value; $H$ is the half channel height.

## IV. RESULTS AND DISCUSSION

### A. Transition over the flat plate

Before analyzing the ribbed plate case, we first examine the effect of the applied wall perturbations on flat plate flow to establish a basis for comparison. This analysis will focus on how these perturbations influence the development of boundary layers, transition to turbulence, and flow stability. To investigate the response of the boundary layer to wall disturbances in the context of laminar-to-turbulent transition, a controlled transition process was initiated using a TS wave at a specified frequency, following the method described in [47]. The disturbances are assumed to be TS waves, with the general form as follows:

$$\psi = \phi(y)exp(i\alpha x - i\omega t) \tag{17}$$

In the above equation, $\psi$ is the stream function for fluctuation velocities. After substituting the above equation into two-dimensional Navier-Stokes equations and mass conservation equation, the Orr-Sommerfeld equation (OSE) can be obtained after some manipulation as

$$(\alpha U - \omega)(\phi'' - \alpha^2 \phi) - \alpha U'' \phi + \frac{i}{Re_{\delta^*}}(\phi'''' - 2\alpha^2 \phi'' + \alpha^4 \phi) = 0 \tag{18}$$

Table 2. Comparisons of eigenvalues for OSE.

| S.No. | Re | $\omega$ | $\alpha$ (Jordinson [48]) | $\alpha$ (Present Result) |
|---|---|---|---|---|
| 1 | 336 | 0.1297 | 0.30864 + i0.00799 | 0.3084 + i0.007985 |
| 2 | 598 | 0.1201 | 0.30801 − i0.00184 | 0.3079 − i0.001858 |
| 3 | 998 | 0.1122 | 0.30870 − i0.00564 | 0.3086 − i0.005708 |

The OSE was solved using the Chebyshev collocation method with Gauss-Lobatto points [48]. In the Eq. 18, the streamwise velocity profile ($U$) is used a Blasius velocity profile. The detailed methodology used in the present work for solving the OSE is presented in the literature [48-49] and validation of our code is presented in Table 2. Figure 3 illustrates the neutral stability curve, plotted as a blue line from the present code. The neutral stability curve ($\alpha_i = 0$, $\alpha_i$ is the imaginary part of the wavenumber, shown in Figure 3), presented along the $x$-axis, is compared with results from similar studies using non-dimensional parameters based on the Reynolds number ($Re_{\delta^*}$) [49]. Following equation (16), the region enclosed by the neutral stability curve represents the area where infinitesimal disturbances are expected to grow exponentially. The red marker in

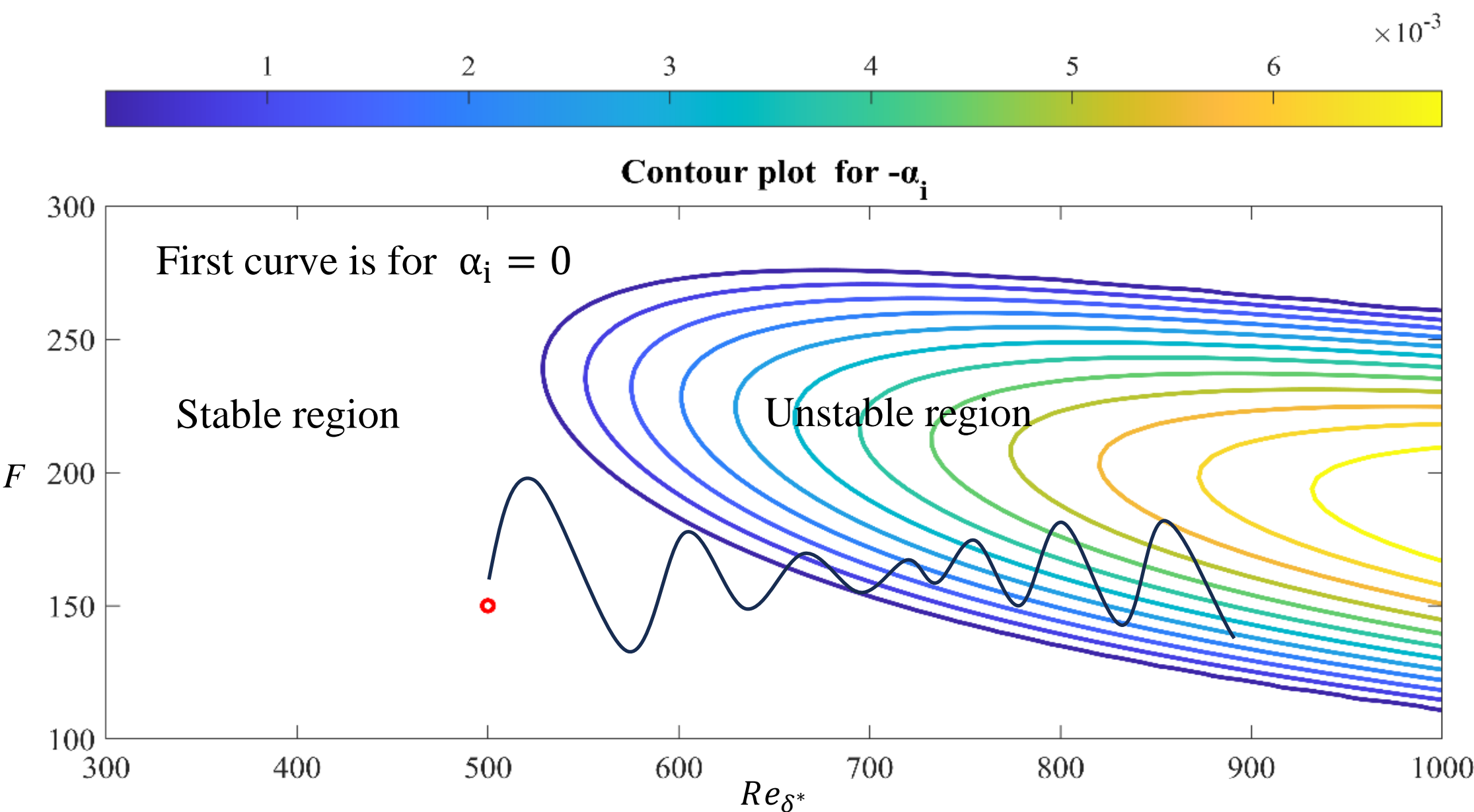

Figure 3. Stability curves for Blasius profile.

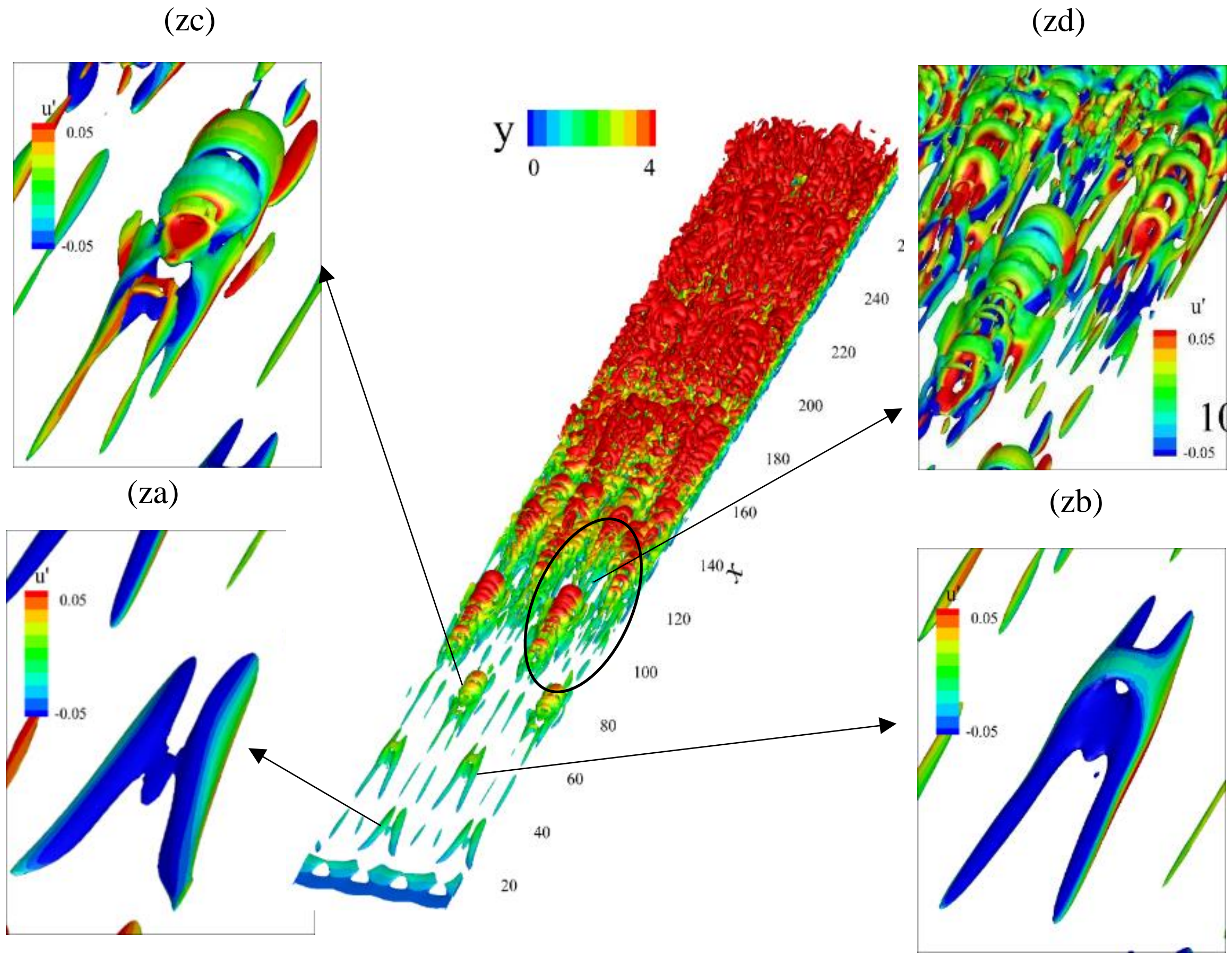

Figure 4. Iso-surfaces of second invariant of the velocity gradient tensor colored by wall normal height. Figures za, zb, zc, and zd are the zoom view of different structure and colored by streamwise fluctuation velocity.

Figure 3 indicates the point for the present computation. For the current simulation, the Reynolds number at the inlet was set to 500, with a non-dimensional wall perturbation frequency ($F = \omega\nu/U_{\infty}^{2}$) of 150. The marker lies within the stable region. By perturbing stable flow, one can study how instabilities develop from initially stable conditions, observing mechanisms such as disturbance amplification, wave formation, and breakdown to turbulence. This helps identify specific modes or instabilities that lead to transition. Initially, a small amplitude wall perturbation ($a_f/U_{\infty}$= 1.5%) was applied, and it was observed that the flow remained stable for the flat plat case, as the disturbances decayed before reaching a critical value of Reynolds number. To induce transition, the amplitude of the wall perturbation was increased to 3%. This larger perturbation also decayed the initial part of the region with space (see Figures 12 (a-c)), but it was sufficient to trigger the flow into a transitional state, leading to the onset of instability and growth of disturbances.

The visualization of instantaneous vortical structures using the Q-criterion [44] provides insight into the development and evolution of turbulence within the boundary layer. By examining these structures, we can better understand how vortices interact, lift, and contribute to the transition and turbulence generation in the flow. Figure 4 illustrates the instantaneous vortical structures inside the boundary layer for the FP case, where iso-surfaces of the second invariant of the velocity gradient tensor, Q, are visualized to highlight the flow's vortical structures. These Q-criterion iso-surfaces (Q = 0.0005) are coloured by the y-coordinate, helping to illustrate the upward movement of flow structures as they develop downstream. Close-up views, labelled za, zb, zc, and zd, show the coherent structures coloured by axial velocity fluctuations, revealing details of the evolving flow. Lambda-shaped (Λ-vortex) structures first appear near the wall between $x = 20$ and $x = 40$ as highlighted in the zoom view za. These structures are staggered downstream (from $x = 20$ to $x = 120$), a characteristic feature of the H-type transition [47]. Around x = 30, two identifiable Λ-shaped vortices mark the onset of secondary instabilities. Here, the spanwise length is double the wavelength of the unstable wave, and each Λ-vortex leg forms a pair of counter-rotating vortical centers containing low-momentum fluid, initiating the ejection mechanism. After forming, the Λ vortex transitions into an elongated ring-shaped vortex with a single head (as seen in $z_b$), followed by a sequence of multiple vortex rings, shown in $z_c$ and $z_d$. Some findings in the literature shows that these Λ -vortices do not self-deform into hairpin vortices; rather, they are formed through Kelvin-Helmholtz (K-H) instability [5]. As the flow moves downstream, alternating injections of high- and low-momentum fluid due to wall perturbations at the TS wave frequency create a structured series of counter-rotating vortex heads, sustained by the amplification of shear-layer disturbances through K-H instability. This process, driven by K-H instability, generates repeating coherent vortical structures, which progress into a packet of hairpin vortices, visible in Figure 4. These hairpin heads deviate from the wall, interacting with streamwise vortices and freestream disturbances that ultimately lead to breakdown. During this phase, the hairpin heads stretch and deform, giving rise to fully developed turbulent regions. To understand the dominant wavenumbers, streamwise velocity ($u(x, 1, z)$ ) data are recorded at each streamwise location and at $y = 1$, capturing variations in the spanwise direction. Applying FFT (Fourier transform ( $\hat{u}(k_z) = \int_0^z u(z) \exp(-ik_z z)\, dz$ ) to this data, we identify dominant wavenumbers corresponding to the location of the primary evolving structures. Figure 5(a) shows that at $x = 10$, the dominant wavenumber reflects the inlet disturbance, whereas further downstream at $x = 21$, a new mode

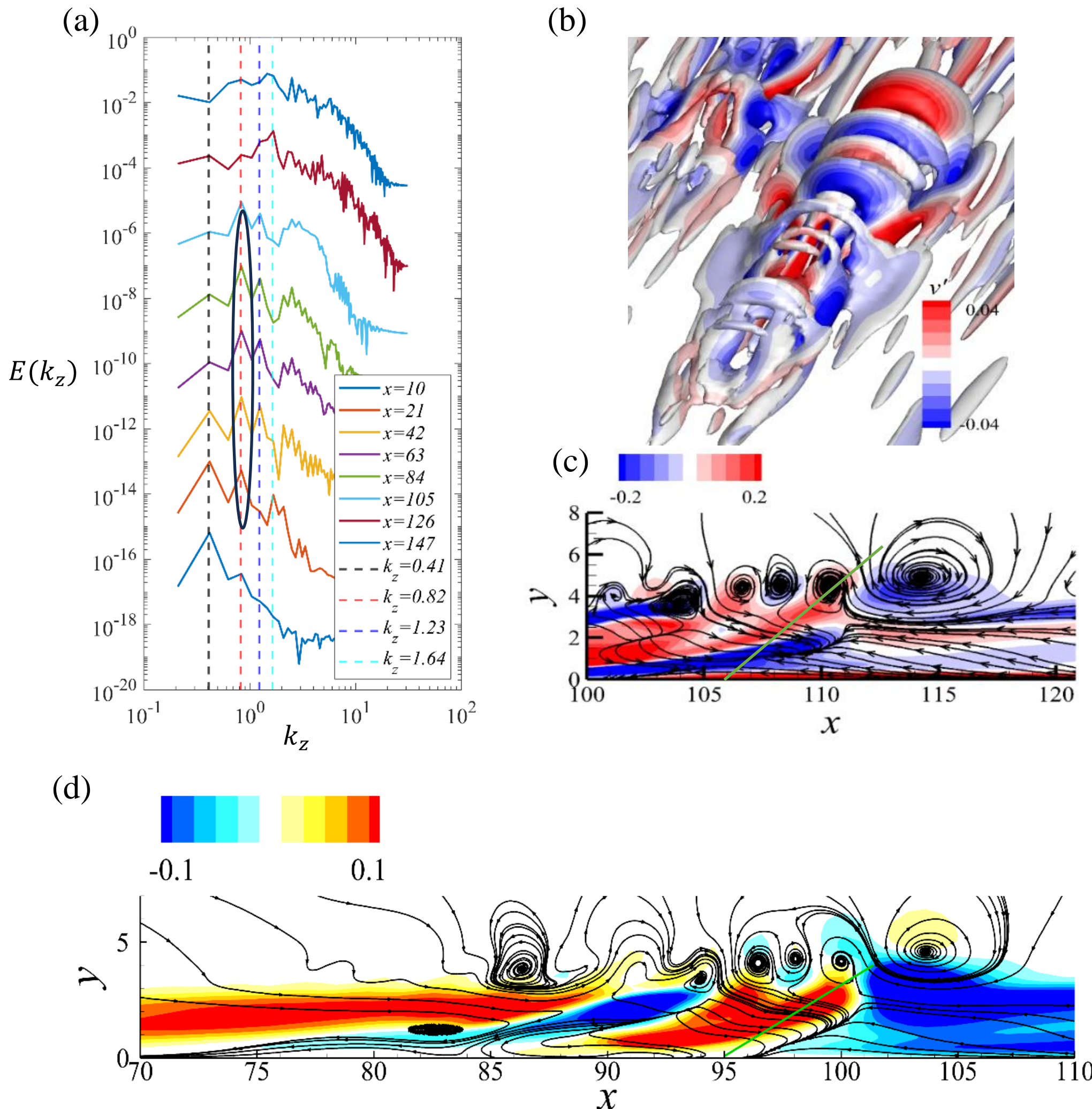

Figure 5 shows the flow features for the FP case; (a) presents the evolution of the dominant wave number in the streamwise direction; (b) illustrates hairpin structures using iso-surfaces of the second invariant of the velocity gradient tensor, which are colored according to the cross-stream velocity fluctuations ($v'$); (c) displays streamlines of velocity fluctuations colored by spanwise vorticity; (d) shows contours of streamwise ($u'$) velocity fluctuations. The inclined green line in the images (c and d) indicates the inclination angle of these coherent structures.

appears, corresponding to the location of lambda-shaped structures shown in Figure 4. By $x = 42$, three peaks emerge corresponding to the location of elongated, ring-like structures. From $x =$ 21 to $x = 105$, the second subharmonic strengthens (marked in Figure 5(a)), reinforcing the lambda structures, which are key in transitioning to turbulence. This dominant wavenumber shows cascade from the dominant wavenumber to emerging subharmonics illustrates the sequential destabilization and energy transfer driving the flow to turbulence. Each stage adds structural complexity, enhancing wave mode interactions and amplifying the instabilities that underpin the turbulent state. In Figure 5(b), the iso-surfaces of $Q$ are colored by streamwise velocity fluctuations ($u'$), revealing a sequence of head of hairpin vortices. Figure 5(c) shows a streamline of fluctuating velocity colored by spanwise vorticity, highlighting the alternating sweep and ejection mechanisms near the vortex heads. Ejection mechanisms are also evident between the vortex legs near the wall, contributing to the formation of alternating rotational structures. The green inclined line in Figures 5(c) and 5(d) indicates the inclination of the head of the hairpin structure, which is found to be approximately 45 degrees.

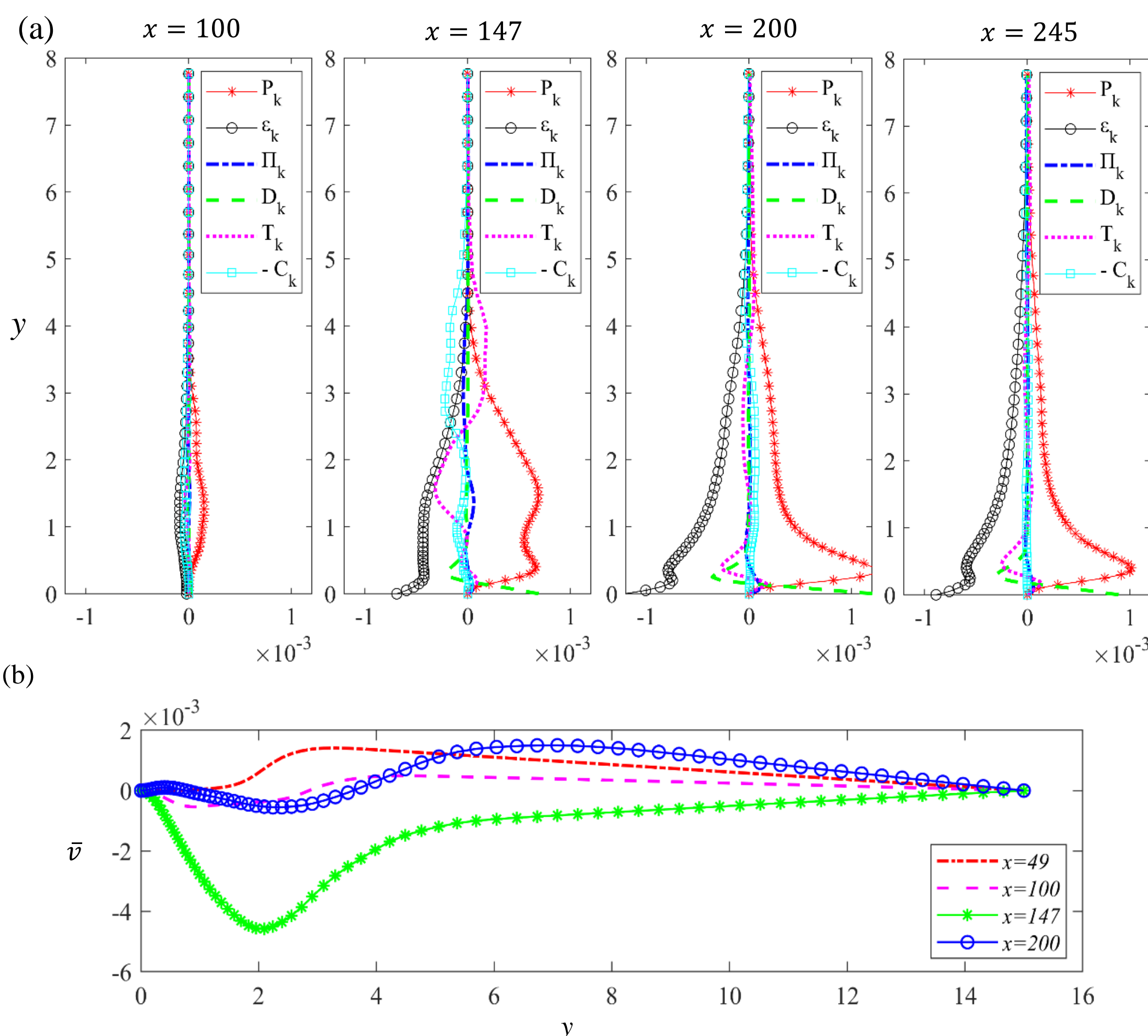

Figure 6. (a) Turbulent kinetic energy budget plots; (b) Evolution of wall normal velocity in streamwise direction.

The analysis of the turbulent kinetic energy (TKE) budget at various streamwise locations highlights the crucial roles of turbulent transport, convection, and production mechanisms in redistributing energy within the boundary layer. This analysis provides insight into how these processes evolve during the laminar-to-turbulent transition and how coherent structures, such as hairpin vortices, contribute to turbulence amplification and energy transfer. The Reynolds-averaged TKE transport equation can be expressed as follows [44]:

$$\frac{\partial k}{\partial t} + C_k = P_k - \varepsilon_k + T_k + D_k + \Pi_k$$

With the convection ($C_k$) = $\overline{u_j} \partial k/\partial x_j$, Production ($P_k$) = $-\overline{\left(u_i' u_j'\right)}(\partial \overline{u_i}/\partial x_j)$, Dissipation ($\varepsilon_k$) = $(1/Re)\ \overline{((\partial u_i'/\partial x_j)(\ \partial u_i'/\partial x_j))}$, Transport ($T_k$) = $(1/2)\left(\partial \overline{u_i' u_i' u_j'}/\partial x_j\right)$, Turbulent Diffusion $D_k$= $(1/Re_\tau)(\partial^2 k/\partial x_j^2)$, Pressure Transport $\Pi_k$=$-\partial \overline{p' u_j'}/\partial x_j$ and, $k = \frac{1}{2}\overline{u_i' u_i'}$.

Figure 6 depicts the spanwise averaged TKE budget across various streamwise locations, showing the spreading of TKE terms in the wall normal direction. The transition begins at $x = 117$ (discussed in the next sections), with the budget plot starting at $x = 100$. Here, the non-linear interaction of structures increases TKE production, whereas energy advection by mean flow and fluctuating velocities becomes more pronounced between $x = 100$ to $x = 147$. As disturbances amplify significantly between $x = 100$ and $x = 147$, TKE production peaks can be seen between $x = 100$ and $x = 147$ due to strong interactions between different scales of motion and the mean flow. At $x = 147$, we observe that the turbulent transport term extracts energy from the near-wall region and redistributes it to the outer region. Up to $x = 100$, due to the absence of free-stream turbulence, no energy transfer occurs from the free stream to the near-wall region. Instead, due to wall perturbation, energy is convected from the near-wall region to the free stream as can confirm from average wall normal velocity (Figures 6 (a) and (b)). As we move downstream, energy accumulates in the free stream. At $x = 147$, where a relatively high magnitude of negative velocity is observed, energy is transported from the free stream to the near-wall region due to the nonlinear interaction of structures in the transition region. In the transition region ($x = 147$), one can observe that turbulent energy transport takes energy from the near-wall region and transfers it back to the free stream. Hence one can say that turbulent transport and convection mechanisms drive this redistribution of TKE energy from the wall to the outer region, increasing TKE across the boundary layer thickness. At $x = 200$, the TKE budget reflects a balance between production and dissipation within the log-law region [44], indicating that this region is close to fully developed turbulence. Closer to the wall, turbulent transport and turbulent diffusion are active, redistributing energy from the near-wall region toward regions farther from the wall. Between $x = 147$ and $x = 200$, high production rates are observed, primarily due to increase in wall shear stress (discussed in next sections) leads to bursts of hairpin vortices that amplify turbulence. This vortex activity is seen in the elevated root mean square (rms) values in Figures 12(b) and (c), underscoring how coherent structures contribute to the surge in turbulence production and energy transfer in this region. Once the flow reaches a fully turbulent state at $x = 245$, turbulent transport mechanisms take prominence near the wall, effectively moving energy from the near-wall region to the outer portions of the boundary layer.

### B. Measurement of drag reduction and transition over the ribbed plate

This study extends the analysis of drag reduction by not only considering the frictional drag reduction over longitudinal riblets in the turbulent region but also examining the overall pressure force loss and surface frictional drag reduction. By evaluating both pressure and shear force losses, we gain a deeper understanding of how riblet geometry influences fluid momentum and turbulence characteristics, providing a comprehensive perspective on the effectiveness of riblets in modifying flow behaviour. To the best of our knowledge, previous investigation only talks about the frictional drag reduction over the longitudinal riblets in turbulent flow region ($DR_{ft}$). In the present study, we will also discuss the overall pressure force loss $PF_oL$ and surface frictional drag reduction $SF_oR$ on the riblet plate as compared to the flat plate. Additionally, this study also discusses the form drag (*F_D*) generated by the leading edge of the roughness elements. We are not merely discussing pressure loss; rather, we are focusing on pressure force loss, as it is directly related to the change in fluid momentum. The pressure force exerted on the fluid, combined with the frictional force acting on the fluid, results in a change in the fluid's momentum. As we know, as the boundary layer develops, loss in pressure is more on the riblets plate as compared to the flat plate. The loss in pressure force in cross plane can be calculated as follows:

$$PF = \int_0^{L_z}\int_0^{L_y}(P_{in}-P_{out})dydz = \frac{1}{L_yL_z}\int_0^{L_z}\int_0^{L_y}\nabla P(y,z)(L_yL_z)dydz = \nabla P_{av}A_{yz} \tag{19}$$

and shear force can be calculated as follows:

$$SF = \int_0^{L_z}\int_0^{L_x}\tau_w(x,z)dxdz = \frac{1}{L_xL_z}\int_0^{L_z}\int_0^{L_x}\tau_w(x,z)(L_xL_z)dxdz = \tau_{w_{av}}A_{xz} \tag{20}$$

Here, $P_{in}$ is the inlet pressure, $P_{out}$ is the outlet pressure, $P_{av}$ is the average pressure in the cross-flow plane, $\tau_{w_{av}}$ is the average wall shear stress over the bottom plate, and $A_{yz} = L_yL_z$ is the cross-sectional area. The above formula can be easily used for the flat plate but for the riblet channel, the shear force formula above cannot be applied directly. For the ribbed plate, the shear force is calculated by calculating the wall shear stress ($\tau_{wn} = 1/Re\ (\partial u/\partial n)$) using the first grid axial velocity and normal distance from the wall and then the wetted area is multiplied.

The overall pressure force loss on the ribbed plate case as compared to the FP case can be calculated as follows:

$$PF_oL\% = \frac{PF_f - PF_r}{PF_f}$$

Similarly, the overall shear force drag reduction on the ribbed plate case as compared to the FP case can be calculated as follows:

$$SF_oR\% = \frac{SF_f - SF_r}{SF_f}$$

Here the subscripts $f$ and $r$ denote the flat and riblet plate cases, respectively.

Due to the presence of the leading edge of the riblets, the form drag percentage is calculated as:

$$F_D\% = \frac{FD_r}{SF_f}$$

Whare $FD_r$ is the form drag on the leading edge of the riblets, calculated from the surface pressure distribution on the cross-section of the riblets. For the flat-plate (FP) case, the value of $F_D$ is zero because no form drag is generated in the absence of riblet leading edges.

Table 3. Details of drag reduction for various longitudinal riblets.

| Nomenclature | $h^+$, $w^+$, $S^+$ | $Cf_tR\%$ | $DR_{ft}\%$ | $PF_oL$ % | $SF_oR\%$ | $F_D\%$ |
|---|---|---|---|---|---|---|
| RP4 | 12.5, 12.5, 12.5 | -58.99 | -6.50 | +10.29 | -13.69 | +3.53% |
| RP3 | 12.5, 25, 25 | -30.58 | -1.00 | +11.23 | -8.80 | +3.51% |
| RP2 | 25, 50, 75 | -6.83 | +28.70 | +49.57 | +22.01 | +7.00% |
| RP1 | 25, 50, 50 | -23.26 | +17.79 | +70.24 | +22.37 | +11.7% |

In Figure 7, the coefficient of friction ($Cf_x = 2\tau_w^*(x)/U_\infty^2$) is plotted for all the computations that are considered. As one can see, the $Cf_x$ plot for the FP case matches well with turbulence correlation ($Cf_x = 0.0594/Re_x^{0.2}$) in the turbulent region. Maximum value of wall-shear stress can be seen between $x = 170$ and $x = 230$, this increase in wall shear stress leads to high amount of production of the kinetic energy. This has been observed in previous section for the flat plat case. As in the $Cf_x$ plot, we can see that in the turbulent region, skin frictional drag reduction is relatively more in the case of RP3 ($h^+ = 12.5$ and $S^+ = 25$) and RP4 ($h^+ = 12.5$ and $S^+ = 12.5$) as compared to the FP case. Notably, in the transition region, the ribbed plate cases exhibit lower wall friction compared to the flat plate. This reduction in friction leads to decreased turbulence TKE production in the ribbed cases which significantly influences the transition process, a phenomenon that will be explored in the following illustrations. In Table 3, the all-reduction parameter is provided for all ribbed plate cases. The positive value shows the increase in parameter, and the negative value shows the decrease in parameter. The parameter $Cf_tR$ represents the reduction in skin friction within turbulent flow regions, whereas $DR_{ft}$ denotes the reduction in skin friction drag force in turbulent flow regions. Additionally, $PF_oR$ and $SF_oR$ indicate the overall reduction in pressure force and shear force, respectively. Additionally, $F_D$ represents the form drag induced by the leading edge of the roughness elements. As one can see, the skin friction coefficient reduction in the turbulent region ($Cf_t$) is seen for all cases. Relatively high $Cf_t$ is seen for the case of RP3(30.58%) and RP4(58.99%). However, reduction in skin friction drag $DR_{ft}$ for RP3 and RP4 is 1% and 6.5%, respectively. Indicating the drag reduction for the riblet size of $h^+$or $S^+ < 25$. This aligns with the previous findings. Choe et al. [11] reported a 6% drag reduction for the $h^+ = 20$ and $S^+ = 17.3$, and for the same riblet size, reported around 4% drag reduction [50, 51]. However, for RP1 ($h^+ = 25$ and $S^+ = 50$) and RP2 ($h^+ = 25$ and $S^+ = 75$), case drag increases by 28.7 and 17.79%, respectively.

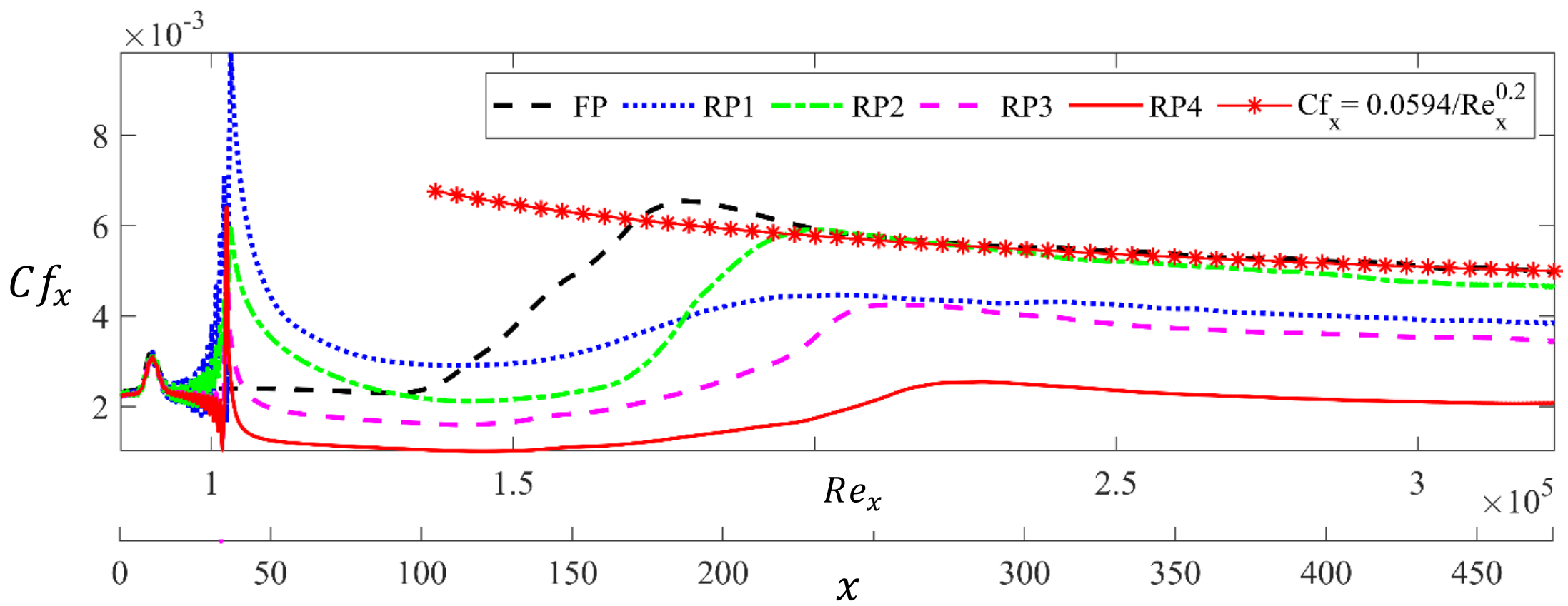

Figure 7. Streamwise variation of spanwise averaged coefficient of friction.

Table 3 indicates that, for the all-ribbed plate, the overall pressure force loss is positive. This indicates that the ribbed plate experiences relatively higher-pressure force losses compared to the flat plate. As roughness height increases the pressure loss increases with a positive value. For the RP4, 10% more overall pressure force loss can be seen as compared to the flat plate. However, for the case of RP1, the overall pressure loss increases by 70% due to an increase in roughness height and width.

In Figure 8, the spanwise average shear force is shown. Although in Figure 7, the skin friction coefficients show less value in the turbulence region for the all-ribbed cases, in Figure 8, one can see that the shear force is relatively high for the turbulent regions for the case of RP1 and RP2 due to the increase in wetted area. These quantities value is shown in Table 3 ($DR_{tf}$) for all cases considered. The overall shear force reduction ($SF_oR$) gives us a glimpse into how different riblet configurations impact wall shear stress. Out of all the cases, RP4 stands out with the most notable reduction at 13.69%, showing the biggest drop in overall-wall shear force. For the case of RP3 shows a decent shear force reduction of 8.80%. On the flip side, RP2 and RP1, which feature larger riblet heights and spacings, show positive $SF_oR$ values of +22.01% and +22.37%, respectively. This suggests that there is an overall increase in wall shear force, likely because the flow interacts more strongly with the bigger riblet structures. Although the local shear force in the turbulent region for RP1 and RP2 shows much variation, the overall shear force drag is almost equal. One important thing that can be noticed is that in the laminar and transition regions, the high riblet spacing shows less increase in drag with same cross-sectional reb geometry (see Figure 7, comparison between RP1 and RP2). The reason behind this will be discuss in upcoming elastration. In the initial part of the laminar-transition region ($x = 50$ to $x = 100$), all riblets case show an increase in the drag force as compared to the flat plate. The RP4 and RP3 cases show significant overall shear drag reduction; however, the form drags generated by the leading edge of the roughness reduces the net drag benefit. For example, in the RP4 case, the shear drag reduction is 13.69%, but when the form drag contribution ($F_D = 3.53\%$) is included, the net drags reduction decreases to approximately 10.16 %. Similarly, for the RP3 case, the net drag reduction is found to be 5.29%. Because, for all the ribbed cases, in the transition regions, the wall shear stress is lower (see Figure 7) compared to the flat plate this indicates that the growth of turbulent structures will be reduced compared to the flat plate, as will be discussed in the upcoming illustrations.

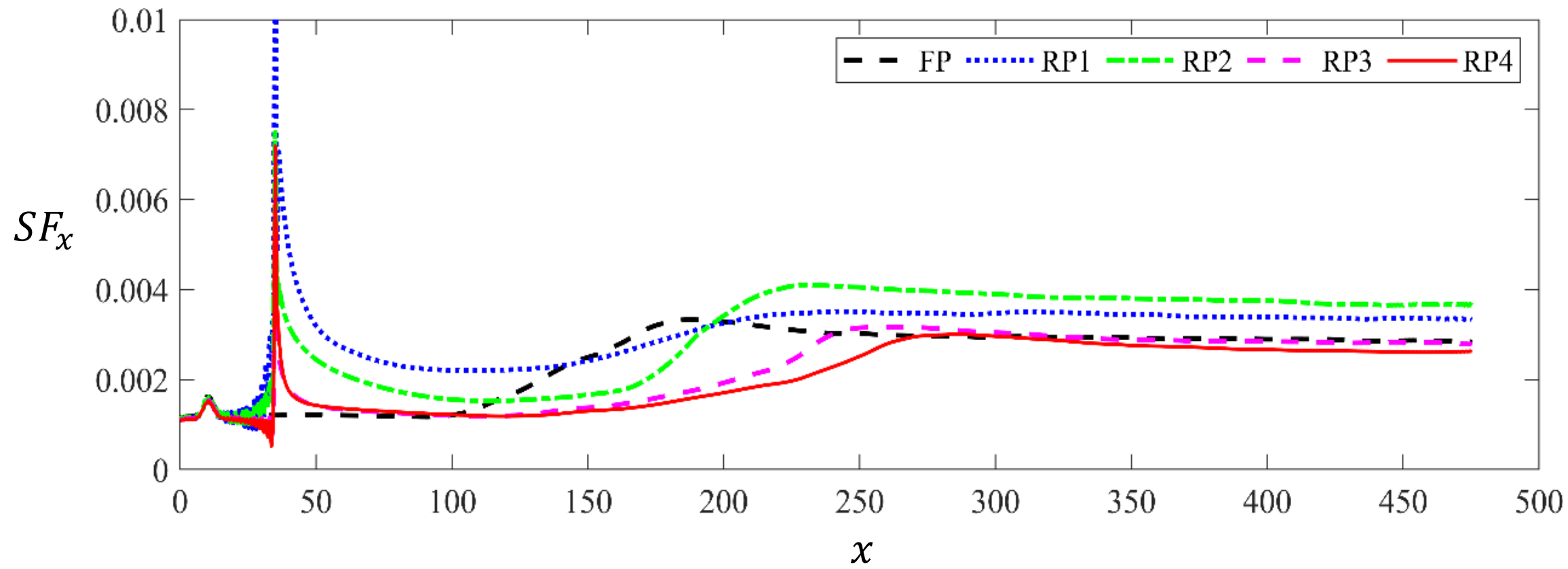

Figure 8. Streamwise variation of spanwise averaged wall shear force.

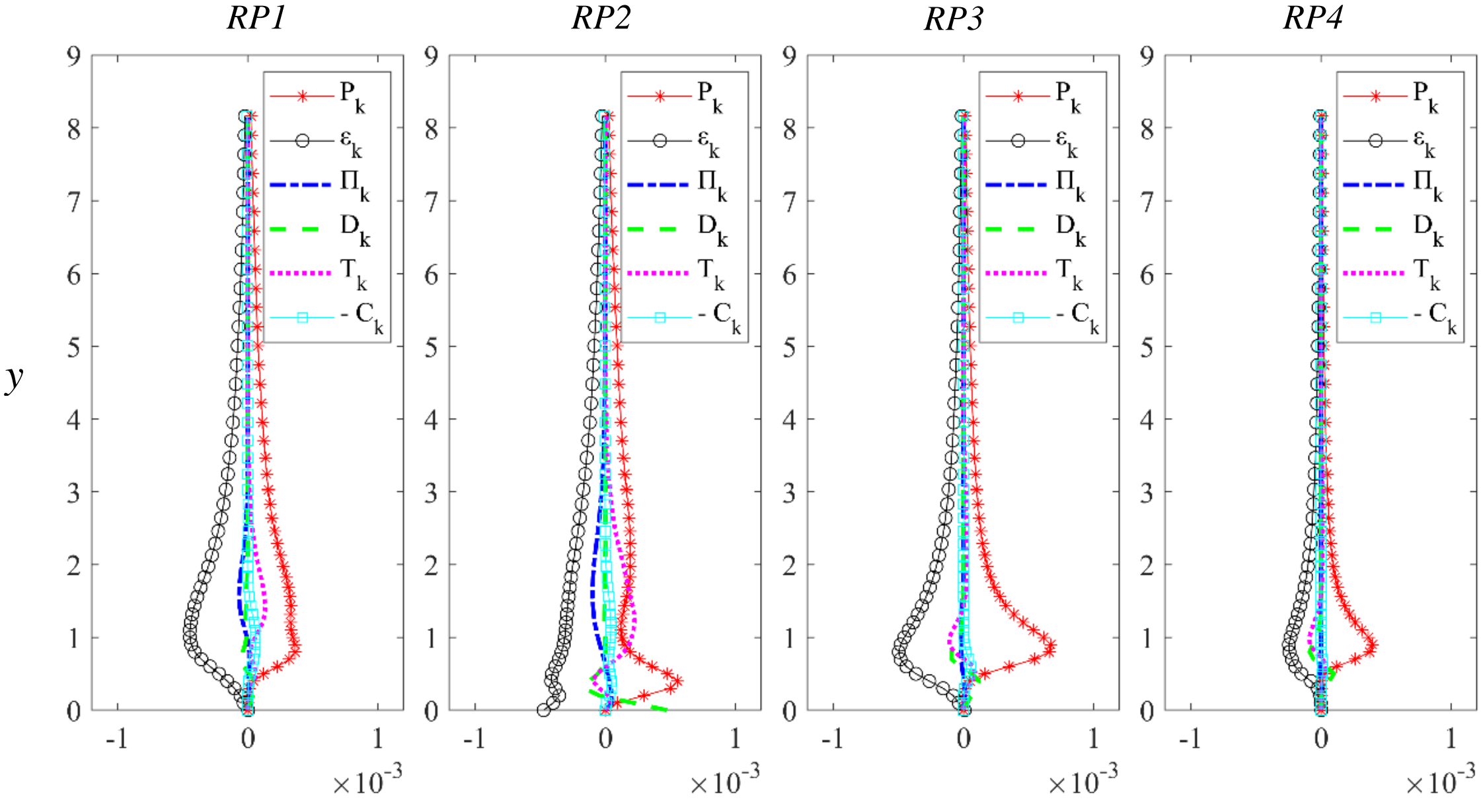

Figure 9. Turbulent budget plot at the mid location between the ribs in fully developed turbulent flow region.

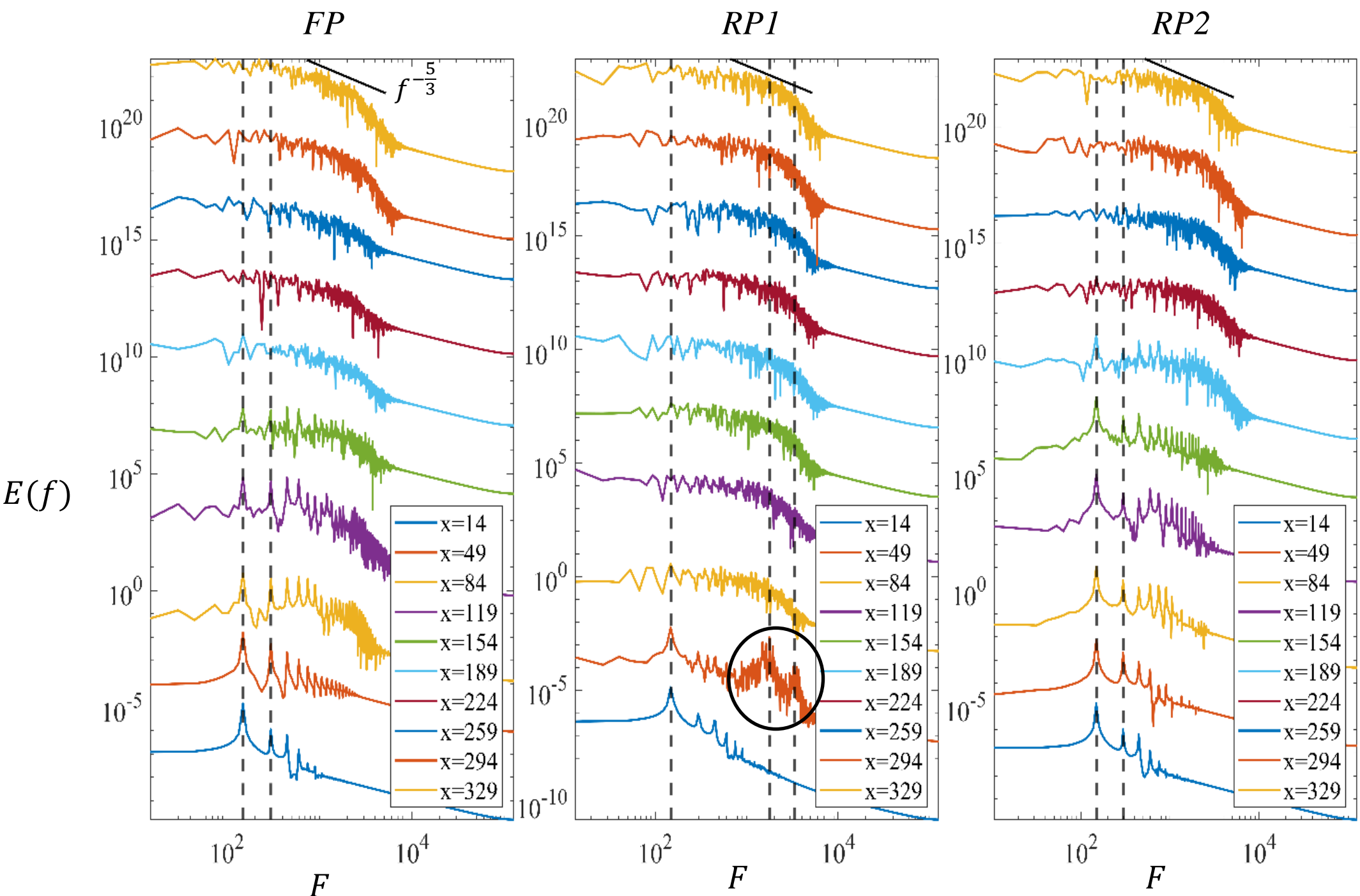

Figure 10. Power spectra of streamwise velocity along the maximum $u_{rms}$ at different $x$ −locations elucidating an amplification of selective frequency.

The variation in near-wall frictional force and the loss in pressure is closely linked to the behavior of near-wall secondary flows and the turbulent transport of momentum toward the wall. This phenomenon will be thoroughly explored in the following illustrations. The TKE budget terms for the ribbed plate case are plotted in Figure 9 at the mid-location between the riblets within the fully developed turbulent flow region. Analyzing these terms provides insight into drag reduction, revealing distinct near-wall behavior influenced by riblet spacing, height, and width. For smaller riblet spacing (RP1, RP3, and RP4), all TKE budget terms approach nearly zero up to a certain distance from the wall, including dissipation of the TKE (as shown in Figure 9). The small-scale structures are closely linked to TKE dissipation [44]. In the ribbed cases, their limited penetration near the wall leads to an outward shift of the dissipation peak. Consequently, the near-wall flow exhibits a small-scale turbulence-free behavior, progressively suppressing turbulence and reducing drag as riblet spacing decreases. In contrast, for larger riblet spacings (RP2), the flow exhibits characteristics similar to a flat plate. The increased spacing generates secondary flows near the wall (as discussed in Figure 17), enhancing turbulence penetration. This leads to higher turbulence levels, as seen in the more pronounced turbulent production near the wall (Figure 9, RP2). The dissipation and turbulent diffusion terms become more significant near the wall, facilitating energy redistribution through turbulent transport, viscous diffusion, and pressure transport mechanisms. In the case of RP1 and RP2, turbulent transport, pressure transport and the convections terms effects are more active compared to RP3 and RP4. For RP3 and RP4 cases with ( $h^+ < S^+ = w^+ = 25$ ), the production peak shifts slightly away from the wall compared to RP1 and RP2. This suggests that optimizing riblet dimensions can effectively suppress near-wall turbulence by limiting quasi-streamwise vortex penetration, reducing energy transfer to the wall, and lowering skin-friction drag. Choi et al. [11] demonstrated that as riblet spacing increases beyond a threshold, streamwise vortices penetrate between the riblets, increasing frictional drag. The typical streamwise vortex diameter in wall units is approximately 25–30 [44], meaning riblet spacings below this range prevent vortex penetration near the wall, causing them to glide above the riblets and reducing drag. This complex interaction between riblet geometry and turbulence mechanisms significantly influences the energy and momentum distribution within the flow, ultimately impacting drag reduction effectiveness.

From Figure 7, the transition onset can be qualitatively predicted based on the wall shear stress plot. To capture the transition onset quantitatively and accurately, contours of spanwise-averaged Reynolds shear stress for all considered cases are presented in Figure 11 to identify the onset of transition. According to the methodology proposed by Ol et al. [52] and Jain and Sarkar [43], the transition onset can be located using a critical Reynolds shear stress value of 0.001. Based on this approach, a delay in the transition onset is evident in the RP2, RP3, and RP4 cases, whare it occurs at $x = 137.5$, $x = 160$, and $x = 172$, respectively. In contrast, the transition onset for the RP1 case is observed at $x = 43$, occurring earlier than in the FP case, where it is observed at $x = 117$. These findings suggest that the riblet plate cases RP2, RP3, and RP4 exhibit a delayed transition compared to the FP case, whereas the RP1 case demonstrates an earlier transition onset. In the case of RP1, the riblets have greater height and width, and the transition occurs earlier compared to other cases. However, when the riblet spacing is increased while keeping the height and width constant (as in RP2), a clear delay in transition is observed. In RP1, the closely spaced roughness elements at the leading edge generate high-frequency horseshoe vortices, as indicated by the circled region in Figure 10 and visualized in Figure 15(a). These vortices arise from the sharp interaction between the flow and the leading-edge geometry, with noticeable high-frequency shedding at $x = 49$ in Figure 10, contributing to the earlier onset of transition. In contrast, RP2 shows an absence of such high-frequency features. Instead, the flow is dominated by streamwise-oriented unsteady vortices, as seen in Figure 15(b), which are less disruptive and lead to a delayed transition. The delay in transition for the roughness cases, relative to the flat-plate case, is determined by comparing the streamwise location of transition onset in each case ($(x_{tr}(flat) - x_{tr}(rough))/x_{tr}(flat) \times 100$, $x_{tr}$ is the transition location for corresponding cases). Based on this definition, the transition delay for the RP2, RP3, and RP4 cases is found to be 17.5%, 37%, and 47%, respectively.

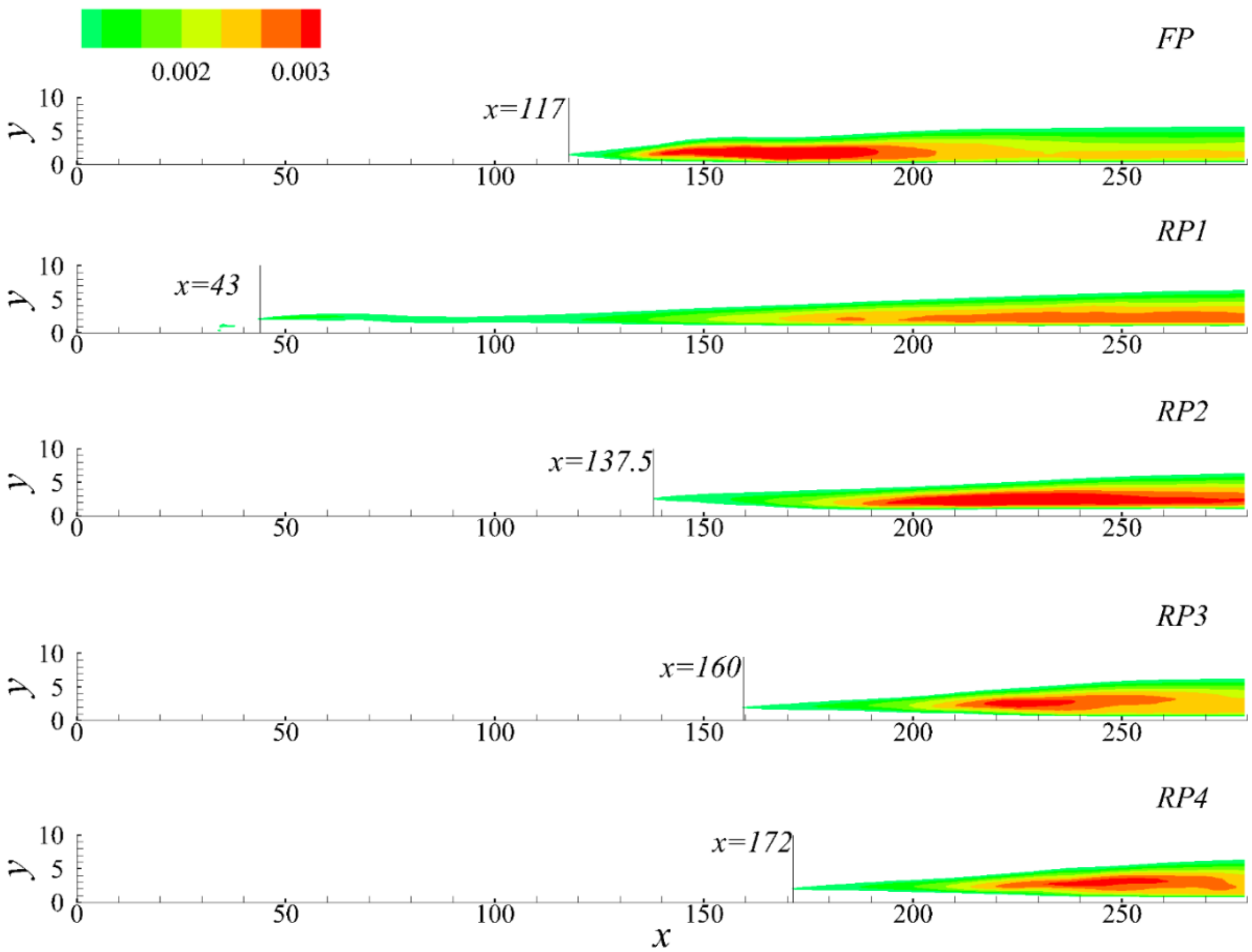

Figure 11. Contour of spanwise averaged Reynolds shear stress ($-\overline{u'v'}$) for all cases considered.

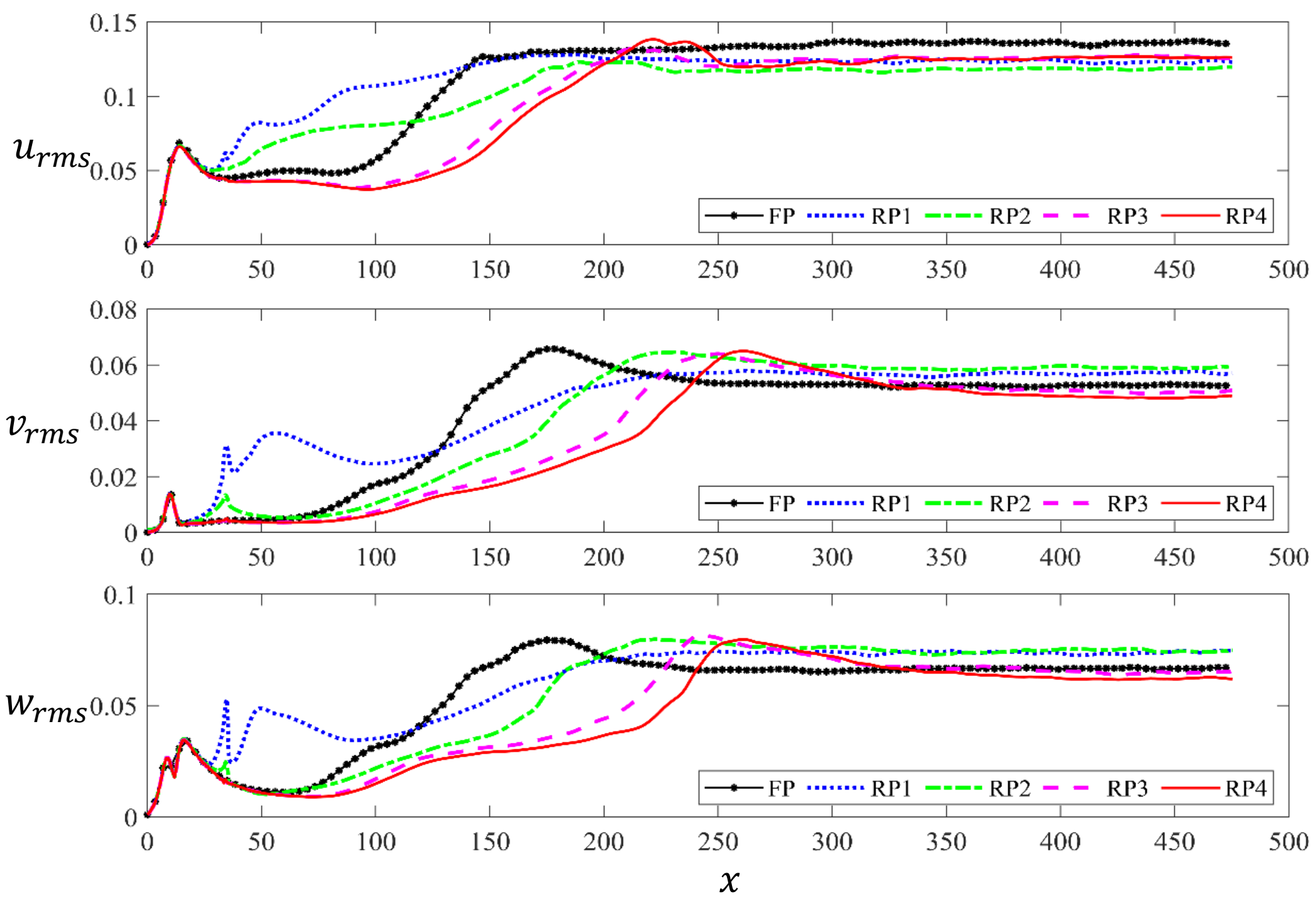

Figure 12. Streamwise evolutions of the maximum rms quantities of velocity fluctuations along the x-direction; (a) $u_{rms}$, (b) $v_{rms}$, and (c) $w_{rms}$.

In Figure 12, the spanwise-averaged maximum rms values of velocity fluctuations are plotted. The figures reveal that the growth rate of the rms values is lower for all ribbed plate cases compared to the flat plate. Although RP1 shows an earlier transition compared to the flat plate, the equilibrium region for fully developed turbulent flow is reached at almost the same location as the flat plate due to the lower growth rate (growth rate is shown in Figure 12). For the other ribbed plate cases (RP2, RP3, and RP4), the transition starts later than on the flat plate, and the lower growth rate further delays the onset of fully developed turbulent flow. Notably, as shown in Figure 7, in the transition regions for all ribbed cases, the wall shear stress is lower, leading to reduced kinetic energy production which slow down the turbulence growth rate compared to the flat plate (Figure 15). In the plot of the rms values of velocity fluctuations, peak values are evident in the transition regions. These peaks are attributed to the abrupt breakdown of hairpin vortices, which intensifies turbulence and leads to elevated fluctuations in these regions.

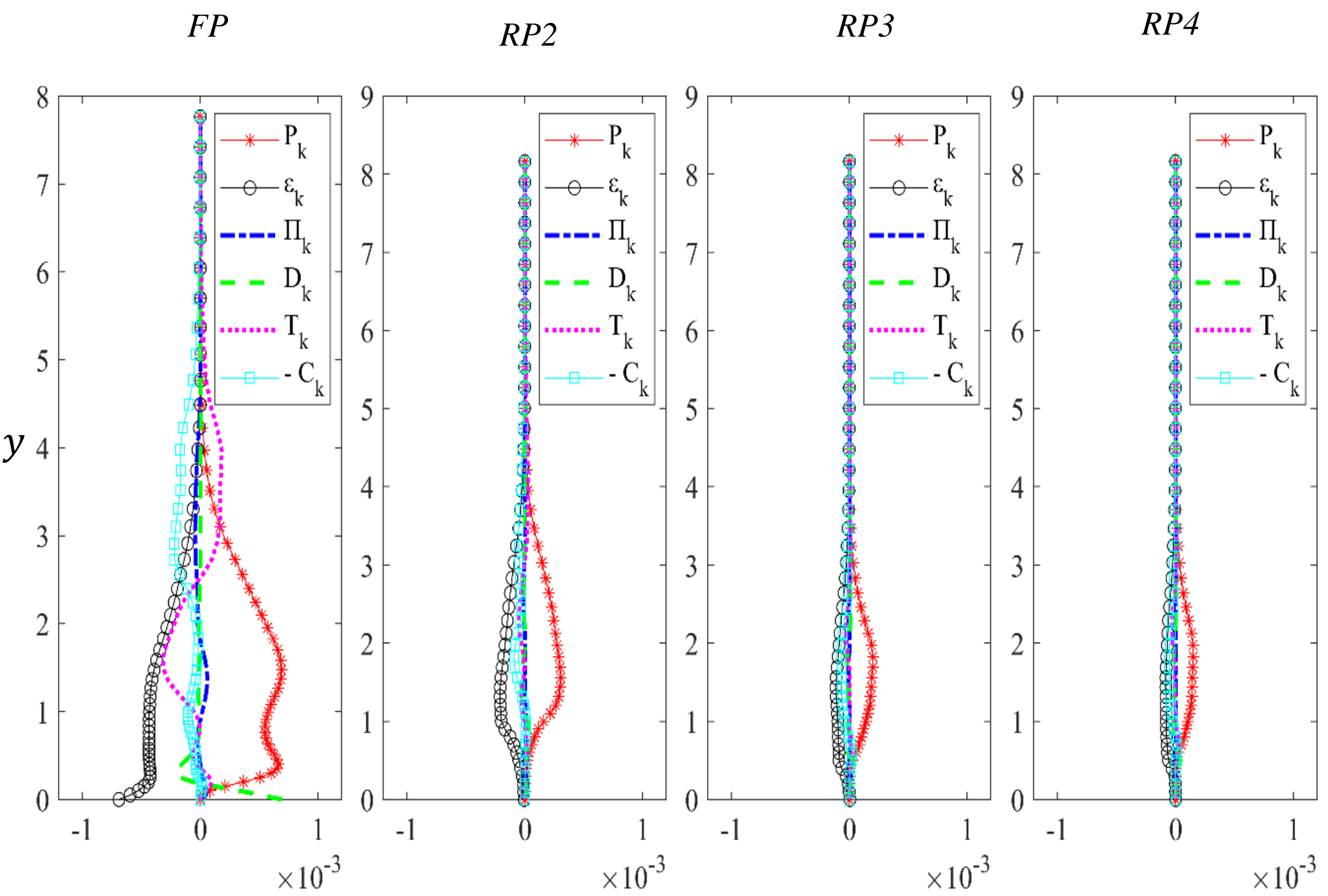

Figure 13. To understand the transition delay mechanism spanwise averaged TKE budget plot at $x = 145$ along the wall normal direction.

In Figure 13, the TKE budget terms are plotted at the same location, $x = 145$, for the flat plate and for the cases which show delayed transition (RP2, RP3, and RP4). For the FP case, higher production of TKE is observed near the wall, where the convection and turbulent transport components of TKE are active. This activity facilitates energy redistribution from the near-wall region to the outer layer. In contrast, in the case with the most delayed transition, TKE production near the wall is significantly lower, leading to reduced convection and turbulent transport. This supporting the slower growth rate of the boundary layer for all the ribbed plate case. Optimized riblet structures can passively reduce drag by altering the flow near the wall, preventing the formation of strong shear layers and thereby limiting turbulent growth.

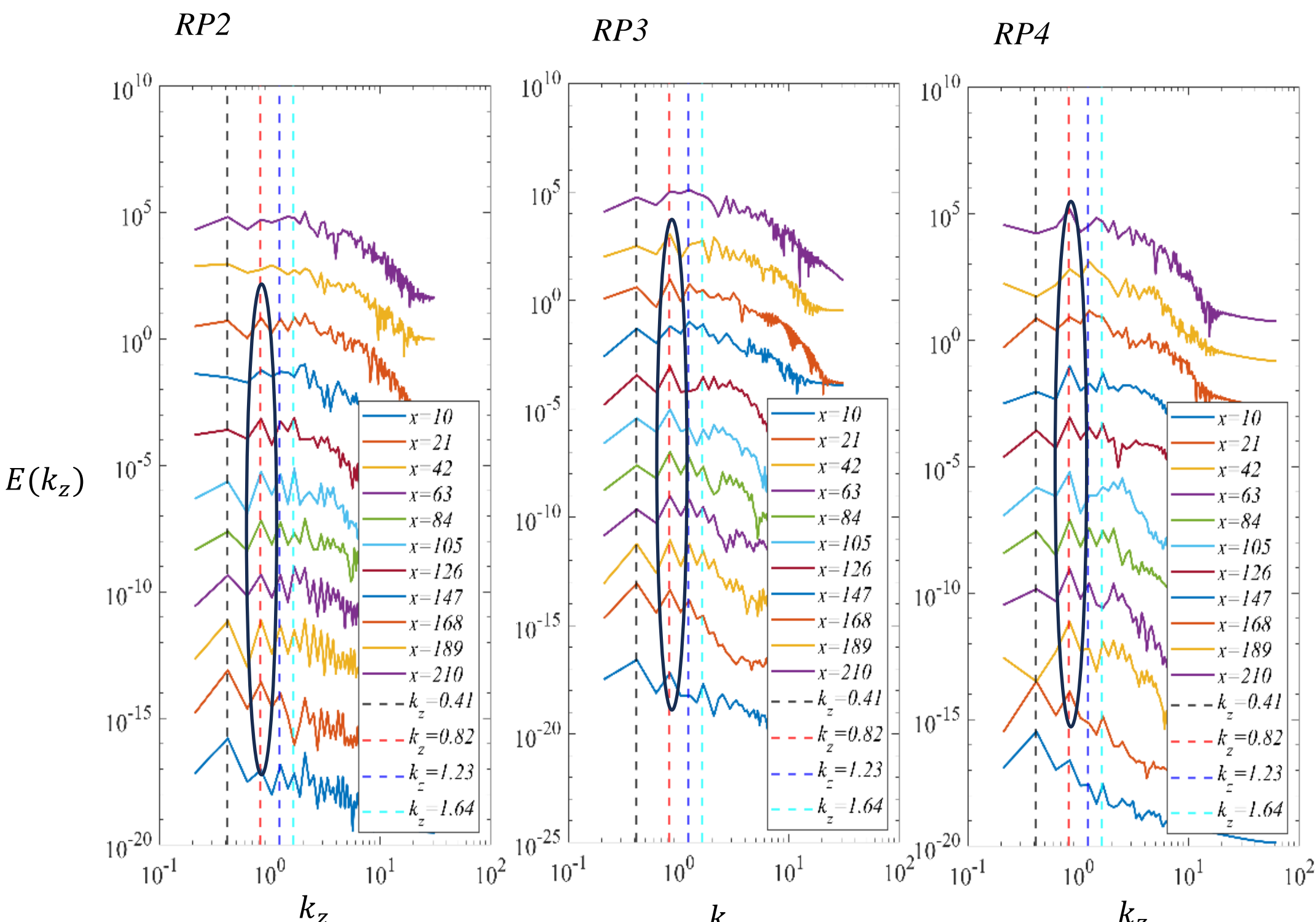

Figure 14. Evolution of the dominant wave number in streamwise direction for ribbed plate cases.

Figure 14 shows how the dominant streamwise wavenumber changes for the ribbed cases. In RP2, one can see multiple wavenumbers at the inlet, with one being the most prominent. As these wavenumbers interact downstream, they help facilitate the shift from laminar to turbulent flow (Figure 15), although this transition happens later than it does for the flat plate (FP) case. On the other hand, RP3 and RP4 only show the dominant wavenumber at the inlet, and their spectral evolution closely resembles that of the FP case. The emergence and eventual fading of a second subharmonic wavenumber, as marked in Figure 14, is particularly noteworthy. This sub-harmonic is closely tied to the development of Λ-shaped structures (which discussed earlier) and is crucial for the transition to turbulence. Interestingly, in all cases except RP1, we see a fully developed flow region once this second sub-harmonic starts to fade (second subharmonic is marked in Figure 14 for RP2, RP3 and RP4). Figure 15 showcases the three-dimensional flow structures for all ribbed cases. For the case of RP1, the larger riblet dimensions create significant disturbances at the leading edge, resulting in high-frequency horseshoe vortices. These vortices highlighted in the zoomed-in view in Figure 15, with their spectral signature clearly shown in Figure 10. For the case of RP2, the wider riblet spacing gets rid of these high-frequency vortices (Figure 10), but the leading edge still causes some moderate disturbance, which weakens the formation of coherent Λ-shaped structures. This leads to a transition process that resembles the FP case, but with a slower development of turbulence. Increase in disturbance at the leading edge for the RP1 case results in high wall shear stress force in the transition region, despite its smaller spacing compared to the RP2 case, as discussed previously. For the case of RP3 and RP4, the smaller riblet dimensions create minimal disturbance at the leading edge. Consequently, well-defined Λ-shaped structures form, and because of the reduced wall shear stress, their growth is slower. This ultimately contributes to a further delay in the onset of transition. Notably, the fully developed turbulent flow region is reached when a single hairpin vortex structure breaks down into multiple smaller structures. This transition behavior is consistently observed for the case of FP, RP2, RP3, and RP4, all of which exhibit almost similar transitional characteristics.

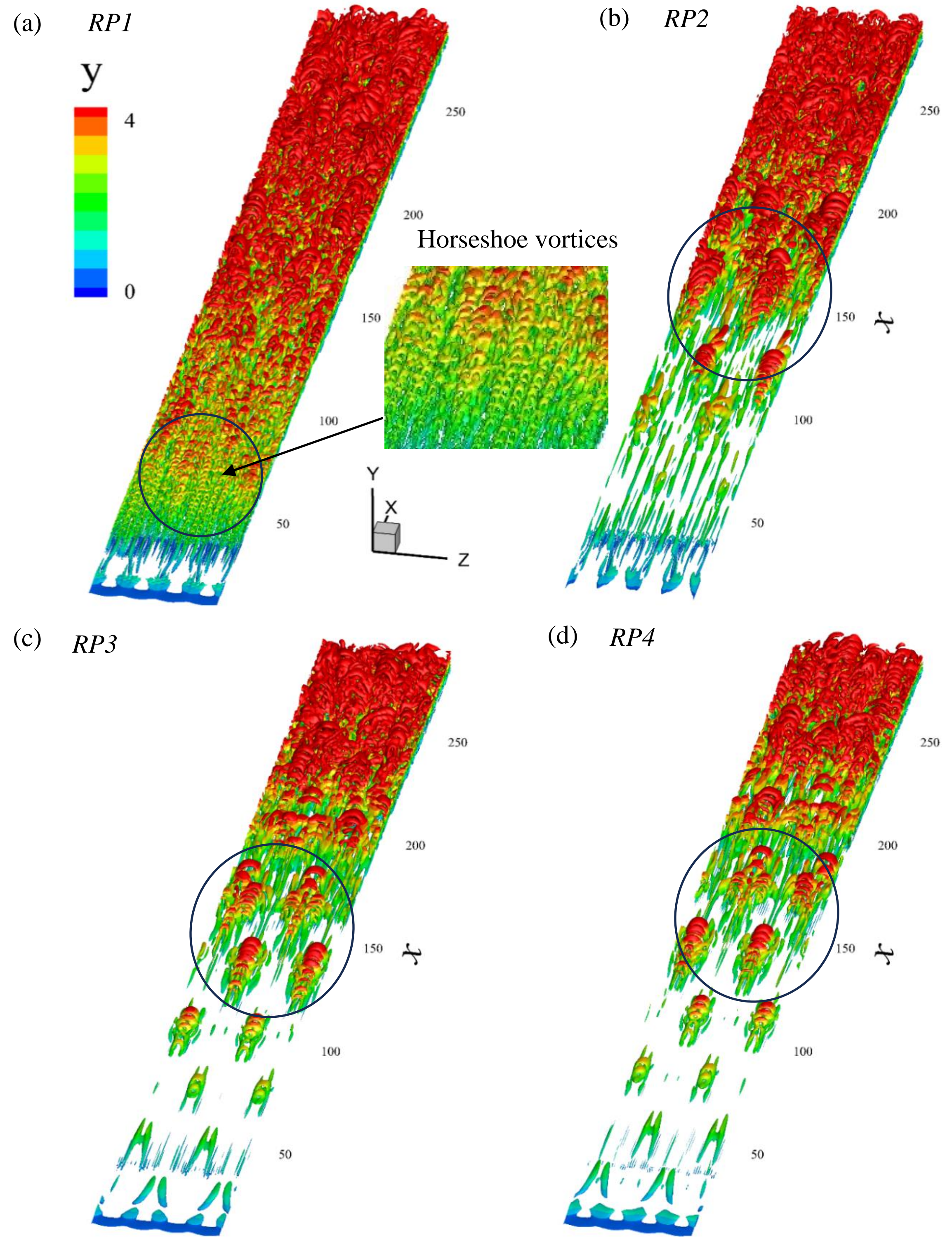

Figure 15. Iso-surfaces of second invariant of the velocity gradient tensor (Q=0.0005) colored by wall normal height.

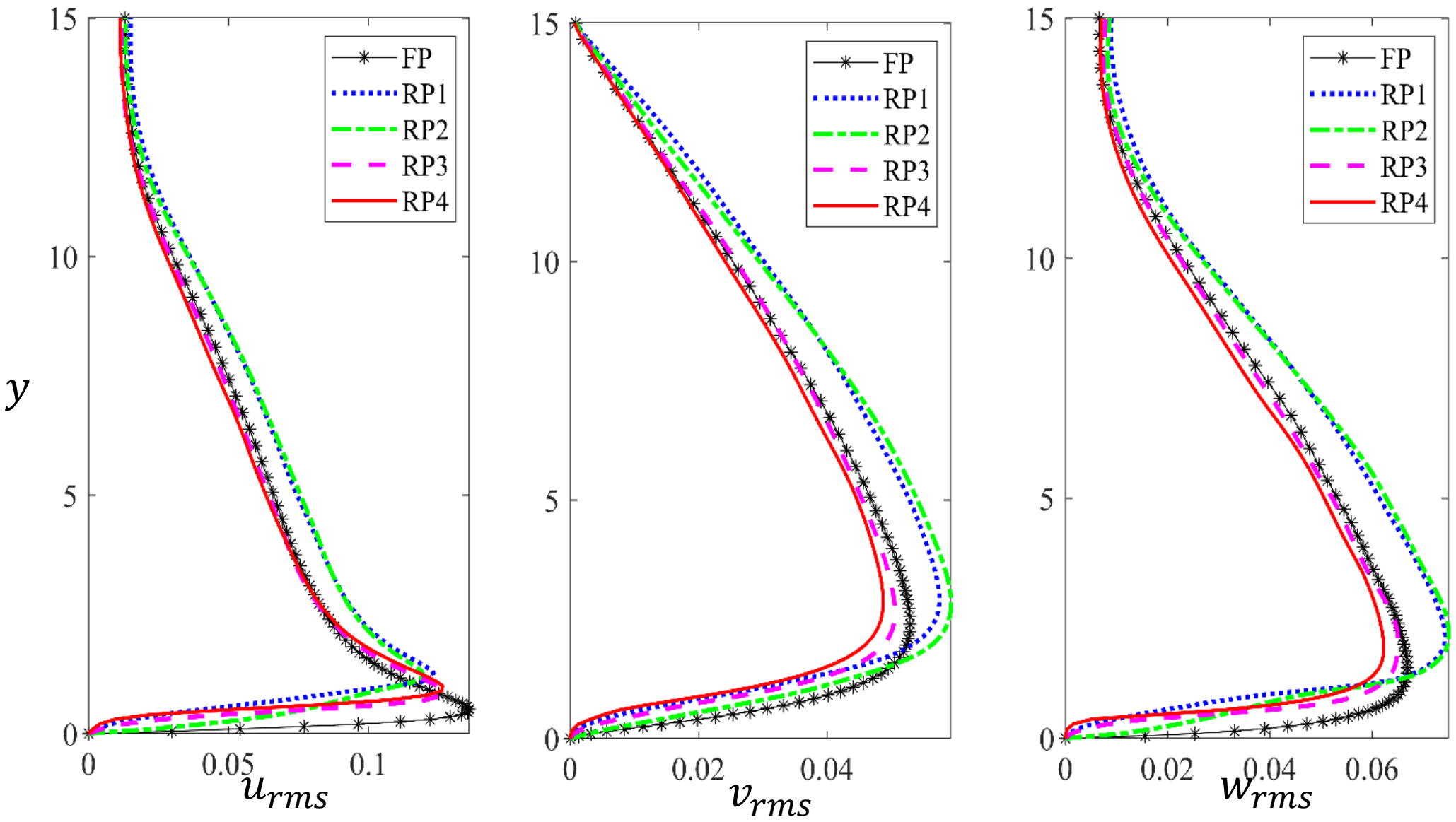

Figure 16. Spanwise averaged (including solid and fluid regions) rms values of velocity fluctuation along wall normal direction in fully developed turbulent flow region.

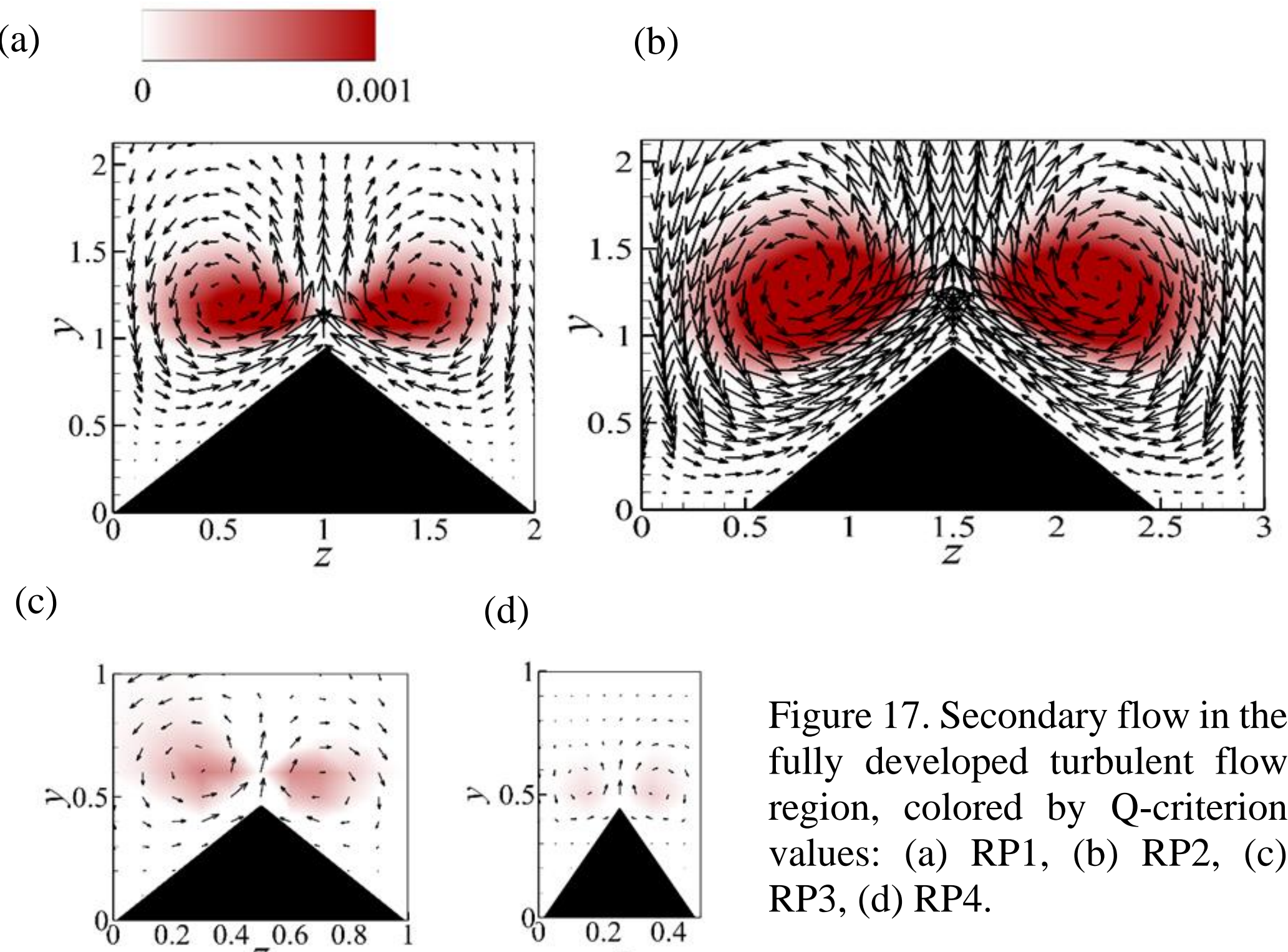

Figure 17. Secondary flow in the fully developed turbulent flow region, colored by Q-criterion values: (a) RP1, (b) RP2, (c) RP3, (d) RP4.

By analyzing the spanwise-averaged rms values of velocity fluctuations and the development of near-wall secondary flows, we gain insight into how different riblet configurations influence turbulence characteristics. In Figure 16, the spanwise-averaged rms values of velocity fluctuations along the wall-normal direction in fully developed turbulent flow regions are presented, encompassing both solid (riblets with zero velocity) and fluid regions [44]. It is observed that for the RP1 and RP2 cases, the rms values above the riblet height are higher compared to the FP case. However, below the riblet height, the opposite trend is observed. For the RP3 and RP4 cases, the rms values are either lower than or nearly equal to those of the FP case. The peak values for the flat plate case are closer to the wall compared to the ribbed cases. In fully developed turbulence, outer layer similarity is observed, indicating that the flow characteristics in this region are similar across different cases. For RP1 and RP2 above the rib almost overlap, confirming this outer layer similarity. Similarly, for RP2 and RP3, the rms values show minimal deviation above the rib height, further suggesting that the outer layer dynamics are largely consistent across these configurations. However, for the cases with different rib heights (e.g. RP1 and RP3), the outer layer similarity is not seen.

In Figure 17, the near-wall secondary flow is depicted in a fully developed turbulent region. The RP2 case exhibits a relatively larger and stronger secondary flow compared to the RP1 case, indicating that increasing riblet spacing enhances both the size and strength of the secondary flow. Maintaining the same riblet aspect ratio but reducing the height (RP3) results in a smaller secondary flow with reduced strength near the wall. Furthermore, when the riblet aspect ratio is reduced (RP4) while keeping the riblet height the same as in RP3, the secondary flow size decreases even further, with negligible strength, as evident from the comparison between the RP3 and RP4 cases. The instantaneous snapshots show the same behavior observed in the averaged secondary motion [11]. In the case of RP4, which exhibits the most delayed transition and significant drag reduction, secondary flows are notably weak and do not reach the wall. However, as the gap between the riblets increases, secondary flows begin to penetrate the near-wall regions. This penetration directs fluid towards the wall with relatively high wall-normal velocity, leading to an accumulation of high-momentum fluid near the wall. Consequently, this leads to an increase in wall shear stress (discussed in the previous elastration). The elevated wall shear stress generates more pronounced TKE near the wall, which ultimately contributes to an increase in drag and an earlier transition to turbulence.

## V. CONCLUSION

This study employed Large Eddy Simulation (LES) to investigate the effects of longitudinal triangular riblets on the laminar-to-turbulent transition in boundary layer flows. Five cases are examined: one with a flat plate and four with ribbed plates. For the FP case, a bidirectional energy transport is observed in the transitional boundary layer, where turbulent convection transport term initially extracts energy from the near-wall region and later redirects it from the free stream back toward the wall, sustaining turbulence growth. Wall perturbations at the TS-wave frequency promote the formation of Λ-shaped coherent structures, which evolve into hairpin structures and interact with the free stream, ultimately leading to the breakdown into fully developed turbulence. The results show that riblets attenuate Tollmien–Schlichting (TS) waves, reducing the growth rate of velocity fluctuations compared to the flat plate for all the cases considered. For the higher riblet height and width case (RP1, $h^+ = 25, w^+ = S^+ = 50$), an early transition occurs due to high-frequency signals (shading of the horseshoe structures) introduced by the leading edge of the riblets. However, increasing the riblet spacing while maintaining the same height and width (RP2, $S^+ = 75$), eliminates these signals, resulting in a transition delay of 17.5%. Both cases exhibit increased drag relative to the flat plate, with RP2 experiencing higher frictional drag due to larger riblet spacing, which promotes the formation of larger vortices. Additionally, RP1 shows higher pressure loss than RP2, indicating that increased riblet spacing reduces fluid pressure loss. When riblet height is reduced (RP3 $h^+ = 12.5$ and $w^+ = S^+ = 25$) while maintaining the same aspect ratio as RP1, the leading edge induces fewer velocity fluctuations, delaying transition by 37% with a modest 8.8% reduction in overall friction drag. The most significant result is observed in (RP4, $h^+ = w^+ = S^+ = 12.5$), where transition is delayed by 47%, and overall friction drag is reduced by 13.69%, with a transition process similar to that of the flat plate. Increase in disturbance at the leading edge for the RP1 case results in high wall shear stress force in the transition region, despite its smaller spacing compared to the RP2 case. For the case of RP4, the shear drag reduction is 13.69%, but when the form drag contribution is included, the net drags reduction decreases to approximately 10.16 %. Similarly, for the RP3 case, the net drag reduction is found to be 5.29%. The study highlights the significant influence of riblet geometry on near-wall turbulence and drag reduction. Smaller riblet spacings (RP1, RP3, and RP4) effectively suppress turbulence by preventing the penetration of streamwise vortices below the rib height, resulting in a near-wall flow that is free from small-scale turbulence (due to near-wall dissipation approaching zero) and leading to reduced skin friction drag. In contrast, larger riblet spacings (RP2) generate stronger secondary flows, enhancing turbulence penetration and increasing drag. In the transition regions, all ribbed cases exhibit lower wall shear stress compared to a flat plate, leading to reduced kinetic energy production which slow down the growth rate of turbulence. Additionally, riblet spacing and dimensions play a crucial role in aerodynamic performance, with smaller riblets minimizing pressure loss and frictional drag. Velocity fluctuations are also affected by riblet geometry, as RP1 and RP2 show higher root mean square (rms) values above the riblets compared to a flat plate, whereas RP3 and RP4 exhibit lower or similar values. In fully developed turbulence, outer layer similarity is observed for the same height of the riblets. The strength of secondary flow depends on riblet spacing, larger spacings (RP2) generate stronger secondary flows, increasing wall shear stress and turbulence, while reduced riblet height and spacing (RP3) or aspect ratio (RP4) weakens secondary flow, contributing to delayed transition and drag reduction. In the flat plate case, dissipation and turbulent diffusion near the wall reach their maximum values, whereas in ribbed cases, these quantities are significantly reduced. Overall, the findings suggest that optimizing riblet geometry-particularly by reducing height, width, and spacing-can delay transition, reduce drag, and enhance aerodynamic efficiency. However, this study did not explore riblet dimensions below 12.5 in wall units, highlighting the need for further investigations to assess their impact on transition to turbulence and drag reduction.

## ACKNOWLEDGEMENTS

The authors express their sincere gratitude to the high-performance computing facilities (HPC and Param Sanganak) of the Indian Institute of Technology Kanpur. One of the authors (GB) acknowledges the support of JC Bose National Fellowship of SERB (DST), Government of India.

## DATA AVAILABILITY STATEMENT

The data that support the findings of this study are available from the corresponding author upon reasonable request.